 \documentclass[trackchanges,twocolumn]{aastex7} 
\usepackage{amsmath}
\usepackage{enumitem}
\usepackage[T1]{fontenc}

\usepackage{mathrsfs}
\usepackage{graphicx}
\usepackage{subcaption}

\begin{document}

\title{Fuzzy Dark Matter and the Impact of Core--Halo Diversity on Its Particle Mass Constraints}

\author[0009-0004-1126-0286]{Dafa Wardana}
\affiliation{Astronomical Institute, Tohoku University, Sendai, Miyagi, 980-0845, Japan}
\email[show]{dafaward@astr.tohoku.ac.jp}  

\author[0000-0002-8758-8139]{Kohei Hayashi}
\affiliation{National Institute of Technology, Sendai College, Sendai, Miyagi, 989-3128, Japan}
\affiliation{Astronomical Institute, Tohoku University, Sendai, Miyagi, 980-0845, Japan}
\affiliation{ICRR, The University of Tokyo, Kashiwa, Chiba, 277-8582, Japan}
\email{khayashi@sendai-nct.ac.jp}

\author[0000-0002-9053-860X]{Masashi Chiba} 
\affiliation{Astronomical Institute, Tohoku University, Sendai, Miyagi, 980-0845, Japan}
\email{chiba@astr.tohoku.ac.jp}

\author[0000-0002-5032-8368]{Elisa G. M. Ferreira}
\affiliation{Kavli IPMU (WPI), UTIAS, The University of Tokyo, Kashiwa, Chiba, 277-8583, Japan}
\email{elisa.ferreira@ipmu.jp}



\begin{abstract}


We investigate how diversity in the core--halo mass relation and the inclusion of higher-order velocity moments affect constraints on the fuzzy dark matter particle mass ($m_\psi$) inferred from the internal kinematics of dwarf galaxies. Using stellar line-of-sight velocities and projected positions for eight Milky Way dwarf spheroidal galaxies, we model their dark matter halos as solitonic cores embedded within outer Navarro--Frenk--White envelopes. We apply both second- and fourth-order Jeans analyses to derive the posterior distribution of $m_\psi$. Our results show that there are two ranges of $m_\psi$ consistent with the observed kinematics: $\log_{10}(m_\psi/\mathrm{eV}) = -19.72^{+0.64}_{-0.56}$, and a narrower low-mass window $\log_{10}(m_\psi/\mathrm{eV}) = -21.81^{+0.39}_{-0.26}$, both within the 68\% credible intervals. The latter becomes prominent only when core--halo diversity is taken into account, which highlights the sensitivity of the inferred fuzzy dark matter particle mass constraints to our understanding of the core–halo relation. Future observations, providing larger stellar samples and more precise kinematic measurements, will be essential for clarifying the allowed parameter space of fuzzy dark matter.
\end{abstract}



\section{Introduction} \label{Sec:Introduction}

The small-scale challenges faced by the standard $\Lambda$-cold dark matter (CDM) paradigm \citep{BullockBoylan-Kolchin2017}, together with the persistent lack of direct detection of CDM particles,\footnote{However, see \citet{Totani2025} for a recent report on a possible dark matter detection.} have motivated renewed interest in alternative dark matter models. Among these, fuzzy dark matter (FDM)—an ultralight bosonic dark matter candidate with a characteristic particle mass of $m_\psi \sim 10^{-22} \, \mathrm{eV}$—has emerged as a compelling scenario \citep{Hu2000}. Owing to its minimal parameter space, in which the particle mass uniquely determines the phenomenology \citep{Schive2014a}, constraining $m_\psi$ is a central task (see, e.g., \citet{Hui2017, Ferreira2021, Hui2021, Eberhardt2025b} for comprehensive reviews).

A defining feature of FDM is its kiloparsec-scale de Broglie wavelength, which has profound consequences for structure formation \citep{Schive2014a, Mocz2017, May2021, Nori2021}. On cosmological scales, wave interference suppresses the growth of density perturbations below a characteristic mass threshold, reducing the abundance of low-mass halos and potentially alleviating the missing satellites problem \citep{Schive2014a}. On galactic scales, the same wave nature gives rise to an effective quantum pressure that counteracts gravitational collapse, preventing the formation of steep central cusps. Instead, FDM halos develop soliton cores (stable, ground-state solutions of the Schrödinger--Poisson equations) characterized by flat central density profiles \citep{Robles2012, Schive2014a}, which provides a natural resolution to the core–cusp problem.

Dwarf spheroidal galaxies (dSphs) and ultra-faint dwarf galaxies (UFDs) in the Milky Way halo are particularly powerful laboratories because they exhibit the highest known mass-to-light ratios of any galaxy type \citep{McConnachie2012, BattagliaNipoti2022}. These extreme mass-to-light ratios indicate strong dark matter domination, which minimizes baryonic complications and enables direct probes of the underlying dark matter potential. Moreover, FDM predicts that soliton cores are larger and dynamically more significant in less massive halos, further enhancing the sensitivity of dwarf galaxies to the FDM particle mass.

Probing the internal mass distribution of individual dwarfs can be performed through stellar kinematics analysis \citep{Hayashi2021, Zoutendijk2021, Goldstein2022}. One commonly used approach employs the relation between the soliton core mass, $M_c$, and the host halo mass, $M_{200}$, first reported by \citet{Schive2014a, Schive2014b} in the form of $M_c \propto M_{200}^{1/3}$. Studies of dwarf galaxy kinematics have widely adopted this relation as a one-to-one mapping for inferring the FDM particle mass.

However, the universality of this core--halo mass relation (CHR) has been investigated in more detail. Independent simulations have reported differing slopes \citep{Mocz2017, Nori2021, Mina2022} or failed to recover a clear scaling altogether \citep{Schwabe2016}. These discrepancies have been attributed to limitations in simulation volume and resolution \citep{May2021}, as well as to the dynamical state of halos, with unrelaxed systems exhibiting systematic deviations from the original relation \citep{Nori2021}. More recent work has demonstrated that FDM halos populate a broad region in the $M_c$--$M_{200}$ plane rather than a narrow relation, reflecting genuine physical diversity driven by variations in merger histories, relaxation states, and cosmological environments \citep{Jowett2022}.

The recognition of this intrinsic diversity has important implications for particle-mass constraints. Constraints derived under the assumption of a unique CHR may be artificially restrictive, potentially excluding values of $m_\psi$ that remain compatible with stellar kinematic data when alternative combinations of $M_c$ and $M_{200}$ are permitted. This motivates a reassessment of FDM particle mass estimates that explicitly incorporates the diversity of the CHR predicted by simulations.

More generally, the bounds inferred from dwarf galaxy kinematics are sensitive not only to the assumed CHR, but also to several methodological choices entering the dynamical analysis. These include, for example, the specific implementation of the Jeans analysis, the treatment of the mass--anisotropy degeneracy, the adopted priors, the choice of dataset, and the assumed form and matching of the full halo density profile. Such choices can significantly affect the ability of the analysis to distinguish a central core from a cusp and may therefore lead to different constraints on $m_\psi$, even when the same dwarf galaxy is analyzed. This broader modeling dependence is reflected in the diversity of bounds reported in the literature (see Figure \ref{fig:fdm_mass_constraints} and references therein).

One important ingredient in this context is the use of higher-order stellar velocity moments. Standard Jeans analyses typically rely on the second-order line-of-sight (LOS) velocity moment (the velocity dispersion), which is known to suffer from the mass--anisotropy degeneracy and can therefore limit the ability of the data to distinguish between cored and cuspy inner density profiles. Including higher-order moments of the LOS velocity distribution, such as the fourth-order moment (kurtosis), provides additional dynamical information that can help alleviate this degeneracy and improve sensitivity to the inner structure of the dark matter halo \citep{Lokas2002,Genina2020,Wardana2025,Banares2026}.

In this work, we examine how allowing for such diversity, together with including fourth-order LOS velocity moments in the Jeans analysis, affects the constraints on the FDM particle mass inferred from the internal kinematics of Milky Way dSphs. By marginalizing over a broad family of CHR consistent with cosmological simulations, we aim to provide a physically motivated estimate of the allowed range of $m_\psi$.\footnote{While this work focuses on the impact of CHR diversity and the inclusion of fourth-order velocity moments in the Jeans analysis, another study by some of the authors, developed in parallel, explores additional assumptions entering the dynamical modeling, such as prior choices and alternative parameterizations of the halo density profile \citep{Horigome2026}. Taken together, these works aim to systematically assess the robustness of the inferred constraints on $m_\psi$ within this class of kinematical analyses.}


This paper is organized as follows.
In Section~\ref{Sec:Model}, we describe the FDM halo model adopted in this work, including the soliton and Navaro--Frenk--White (NFW) density profile and scaling relations in FDM cosmology.
Section~\ref{Sec:DataAnalysis} presents the dwarf galaxy sample, the kinematic data, and the analysis methodology.
In Section~\ref{Sec:Results}, we present the resulting constraints on the FDM particle mass.
Section~\ref{Sec:Discussion} discusses these results and compares them with constraints from other independent probes.
Finally, Section~\ref{Sec:Conclusions} summarizes our conclusions.


\section{The Model} \label{Sec:Model}
Under the assumptions of dynamical equilibrium and spherical symmetry, the motion of stars in a gravitational potential $\Phi(r)$ is fully described by the phase-space distribution function $f(\mathbf{r}, \mathbf{v})$. In practice, however, the distribution function is not directly observable. Instead, the Jeans equations provide a means to relate the underlying gravitational potential to observable stellar kinematics through moments of the distribution function.

In spherical coordinates $(r, \theta, \phi)$, the velocity moments are defined as
\begin{equation}
\nu \overline{v_r^i v_\theta^j v_\phi^k} = \int \mathrm{d}^3 v \, v_r^i v_\theta^j v_\phi^k \, f,
\label{eqn:moments}
\end{equation}
where $\nu(r)$ is the three-dimensional stellar density distribution and $r$ denotes the radial distance from the system center. The second-order Jeans equation then takes the form \citep{BinneyTremaine2008}
\begin{equation}
\frac{\mathrm{d}}{\mathrm{d} r} \left( \nu \sigma_r^2 \right) + 2 \frac{\beta}{r} \nu \sigma_r^2 + \nu \frac{\mathrm{d} \Phi}{\mathrm{d} r} = 0,
\label{eqn:Jeans2nd}
\end{equation}
with $\sigma_r(r)$ denoting the radial velocity dispersion. Under spherical symmetry, $\sigma_\theta = \sigma_\phi$, which allows the stellar velocity anisotropy to be defined as $\beta(r) \equiv 1 - \sigma_\theta^2(r)/\sigma_r^2(r)$.

Solving Equation (\ref{eqn:Jeans2nd}) for $\sigma_r(r)$ under the assumption of a constant velocity anisotropy, $\beta(r) = \beta$ for simplicity, and projecting the solution along the LOS yields the observable LOS velocity dispersion
\begin{equation}
    \sigma^2_{\rm los}(R) = \frac{2}{I(R)} \int_R^\infty \mathrm{d} r \left( 1- \beta \frac{R^2}{r^2} \right) \frac{\nu \sigma_r^2 r}{\sqrt{r^2 -R^2}},
    \label{sigmalos0}
\end{equation}
in which $R$ is the projected radius, $I(R)$ is the projected stellar density profile derived from $\nu(r)$, and $\sigma_{\rm los}(R)$ is the LOS velocity dispersion. Throughout this work, we adopt a Plummer profile for the stellar density \citep{Plummer1911}, given by $I(R) = (\pi r_h^2)^{-1} [1 + R^2/r_h^2]^{-2}$, where $r_h$ is the projected half-light radius.

Due to projection along the LOS, LOS velocity distributions are generally non-Gaussian \citep{BinneyMerrifield1998}. Therefore, describing LOS velocity distributions solely by their velocity dispersion amounts to approximating a non-Gaussian distribution by its closest Gaussian, which can introduce biases in dynamical inferences (see \citet{Read2021} for a demonstration). Higher-order velocity moments provide a means to quantify deviations from Gaussianity, with increasing moment order encoding progressively "fine-grained" details of the LOS velocity distribution shape. Because dwarf spheroidal galaxies are primarily dispersion-supported systems, we only incorporate symmetric deviations from Gaussianity and characterize them using the fourth-order velocity moments.

The fourth-order Jeans equation for a spherically symmetric system is given by
\begin{equation}
    \frac{\mathrm{d}}{\mathrm{d} r} \nu \overline{v_r^4} + \frac{2\beta}{r} \nu \overline{v_r^4} + 3\nu \sigma_r^2 \frac{\mathrm{d} \Phi}{\mathrm{d} r}=0,
\label{eqn:Jeans4th}
\end{equation}
where $\overline{v_r^4}$ stands for the fourth-order velocity moment. Projecting the solution of Equation (\ref{eqn:Jeans4th}) along the LOS under the assumption of constant $\beta$ yields \citep{Lokas2002, Battaglia2013}
\begin{equation}
\begin{split}
    \overline{ v_{\rm los}^4}(R) &= \frac{2}{I(R)} \int_R^\infty \mathrm{d} r \\
    & \times \left[1-2\beta \frac{R^2}{r^2} +\frac{1}{2}\beta (1+\beta)\frac{R^4}{r^4} \right] \frac{\nu \overline{v_r^4}r}{\sqrt{r^2-R^2}}.
\end{split}
\label{eqn:battaglia2013}
\end{equation}
For convenience, the fourth-order LOS velocity moments are expressed in terms of the LOS kurtosis,
\begin{equation}
    \kappa_{\rm los}(R) = \frac{\overline{v^4_{\rm los}}(R)}{\sigma_{\rm los}^4(R)}
    \label{eqn:kappa}
\end{equation}
\citep{Merrifield1990,Lokas2002}. A Gaussian velocity distribution corresponds to $\kappa_{\rm los} = 3$, while $\kappa_{\rm los} < 3$ and $\kappa_{\rm los} > 3$ indicate distributions with thinner and heavier tails than a Gaussian, respectively.

Fourth-order velocity moments have been shown to be effective in mitigating the mass-–anisotropy degeneracy, primarily because of their sensitivity to the velocity anisotropy parameter \citep{Battaglia2013, Genina2020, Wardana2025, Banares2026}. This degeneracy persists in kinematic analyses within the FDM framework \citep{GonzalezMorales2017, Goldstein2022}, even though the model assumes an intrinsically cored density profile. In the FDM context, the mass-–anisotropy degeneracy manifests as a degeneracy between the velocity anisotropy and the core radius $(r_c)$, which directly impacts the inferred FDM particle mass since $r_c$ is tightly linked to $m_\psi$ through the scaling relations discussed in Section \ref{sec:scaling_relation}. Incorporating fourth-order velocity moments is therefore expected to help exclude regions of parameter space in which acceptable fits to the velocity dispersion are achieved only at the cost of implausible velocity anisotropy. In this work, we incorporate fourth-order velocity moments built upon the dynamical model described in \citet{Wardana2025}, employing uniform and Laplacian kernels to introduce flexibility in the LOS velocity distribution shape. The use of such kernels was originally introduced by \citet{SandersEvans2020}.

\subsection{Dark Matter Density Profiles}
FDM halos naturally develop a central soliton core as the ground-state solution of the Schrödinger-–Poisson equation. Based on numerical simulations, \citet{Schive2014a} provided an empirical form for the density profile of FDM halos,
\begin{equation}
    \rho_{\rm sol}(r) = \rho_c \left[ 1 + 0.091 \left( \frac{r}{r_c} \right)^2 \right]^{-8},
    \label{eqn:solitonprofile}
\end{equation}
where the core radius $r_c$ is defined as the radius at which the density drops to one-half of its central value $\rho_c$. The corresponding core density is given by
\begin{equation}
    \rho_c \approx  1.9 \times 10^{12} \left( \frac{m_\psi}{10^{-23}\mathrm{~eV}} \right)^{-2} \left( \frac{r_c}{\rm pc} \right)^{-4}~\mathrm{M_\odot}~\mathrm{pc^{-3}}.
\end{equation}
Equation (\ref{eqn:solitonprofile}) provides an accurate description of the density profile out to radii of a few $r_c$ \citep{Schive2014b, Jowett2022}. At larger radii, the density departs from the soliton solution and transitions to an NFW-like profile, which can be approximated as
\begin{equation}
    \rho(r) = \rho_s \left( \frac{r}{r_s} \right)^{-1} \left( 1 + \frac{r}{r_s} \right)^{-2},
\end{equation}
where $\rho_s$ and $r_s$ denote the characteristic NFW density and scale radius, respectively.

Accordingly, the halo density profile is modeled as a two-component structure. Defining $r_t$ as the transition radius, the inner region ($r < r_t$) follows the soliton profile, while the outer region ($r \geqslant r_t$) follows the NFW profile \citep{Navarro1997}. In this work, the two components are related by imposing continuity of the density at the transition radius,
\begin{equation}
\begin{split}
    \rho_c \epsilon &\equiv \rho_c \left[ 1 + 0.091 \left( \frac{r_t}{r_c} \right)^2 \right]^{-8} \\
    &= \rho_s \left( \frac{r_t}{r_s} \right)^{-1} \left( 1 + \frac{r_t}{r_s} \right)^{-2}.
\end{split}
\end{equation}
With $\epsilon$ and $r_s$ specified, the scale density $\rho_s$ is uniquely determined, and $r_t$ is given by $ r_t = 0.091^{-1/2} \left(\epsilon^{-1/8}-1\right)^{1/2} r_c$.

In this work, the soliton and NFW components are matched by imposing continuity of the density at the transition radius, without requiring continuity of the density derivatives. While this prescription is commonly used and provides a flexible parameterization for the transition region, we note that enforcing a smoother matching could modify the detailed shape of the density profile near the transition and may, in principle, affect the inferred constraints on $m_\psi$ (S. Ando et al. in prepatarion).

\subsection{Scaling relations in FDM cosmology}
\label{sec:scaling_relation}
Numerical simulations of FDM consistently predict a well-defined relation between the soliton core mass $M_c$ and its core radius $r_c$ \citep{Schive2014b},
\begin{equation}
    a^{1/2} M_c \approx \frac{5.5 \times 10^{12}}{(m_\psi/10^{-23} \mathrm{~eV})^2 (a^{1/2}~r_c/\mathrm{pc})}~\mathrm{M_\odot},
    \label{eqn:Mc-rc_relation}
\end{equation}
where $a = 1/(1+z)$ is the cosmological scale factor, and $z$ is the redshift. This relation implies that more massive soliton cores are correspondingly more compact, reflecting the balance between gravity and quantum pressure, the effective pressure generated by spatial gradients of the ultralight dark matter field, which counteracts gravitational collapse and suppresses structure below the de Broglie scale.

In addition to the core mass--core radius scaling, simulations have identified a correlation between the soliton core mass and the host halo mass, commonly expressed as $M_c \propto M_{200}^{\eta}$ \citep{Schive2014b, Mocz2017, Nori2021}. While the existence of such a correlation appears robust, its slope and normalization exhibit substantial scatter across simulations. To capture this intrinsic diversity, we adopt the generalized CHR proposed by \citet{Jowett2022},
\begin{equation}
\begin{split}
    a^{1/2} M_{\rm c} &= \xi  \left( \frac{m_\psi}{8\times10^{-23}\:{\rm eV}} \right)^{-3/2} \\
    & + \left( \sqrt{\frac{\zeta (z)}{\zeta (0)}} \frac{M_{200}}{\mu} \right)^{\eta}
    \left( \frac{m_\psi}{8\times10^{-23}\:{\rm eV}} \right)^{\frac{3(\eta-1)}{2}} \: {\rm M_{\odot}},
\end{split}
\label{eqn:core_halo_mass_relation_jowett}
\end{equation}
where $M_{200}$ is the halo mass enclosed within the radius at which the mean density equals 200 times the critical density of the Universe. The parameters $(\xi, \mu, \eta)$ encode the normalization, characteristic mass scale, and slope of the relation, respectively. Best-fit values derived from large-volume cosmological simulations are $\xi = 8.00^{+0.52}_{-6.00} \times 10^6~\mathrm{M_\odot}$, $\log_{10}(\mu/{\rm M_\odot}) = -5.73^{+2.38}_{-8.38}$, and $\eta = 0.515^{+0.130}_{-0.189}$, with the quoted uncertainties reflecting the intrinsic scatter among simulated halos \citep{Jowett2022}.

To incorporate this diversity into our analysis, we draw $N_{\textrm{CHR}}$ independent realizations of the parameter set $(\xi, \mu, \eta)$, generating a corresponding ensemble of CHRs that uniformly populate the $\log_{10}(M_c)$--$\log_{10}(M_{200})$ plane within the scatter reported by \citet{Jowett2022}.
This procedure effectively induces a uniform prior on $M_c$.
In this work, we adopt $N_{\textrm{CHR}} = 500$ to ensure a smooth and minimally discrete sampling of the relation space.
Each CHR independently enters the inference through estimation of the second- and fourth-order LOS velocity moments for a given set of model parameters. The estimated moments are then fitted to the data.
Although a fully probabilistic treatment of the core mass--halo mass distribution would be more realistic, the limited number of simulated halos and their incomplete coverage motivate this approach, which nevertheless allows us to quantify the impact of CHR diversity on constraints of $m_\psi$.

\section{Data and Analysis}
\label{Sec:DataAnalysis}
\renewcommand{\arraystretch}{1}
\begin{table*}[htbp]
  \centering
  \setlength{\tabcolsep}{8pt}
  \caption{Observational Properties of the Classical MW's dSphs}
  \label{Tab:observational_data}
  \begin{tabular}{lcccccccc}
    \hline \hline
     Galaxy Name & R.A.         & Decl.        & $N_{\rm star}$ & $M_*$                      & $D_\odot$ & $r_h$ & $v_{\rm sys}$ & References \\
                 & (hh:mm:ss) & (dd:mm:ss) &                & ($10^6$ M$_\odot$) & (kpc)     & (pc)  & (km s$^{-1}$) &      \\
    \hline
    Carina    & 06:41:36.7 & $-$50:57:58 & 1086 & 0.38 & 106 $\pm$ 6  & 308 $\pm$ 23 & $+$220.7 & (1), (4), (11) \\
    Draco     & 17:20:12.4 & $+$57:54:55 & 468  & 0.29 & 76 $\pm$ 6   & 214 $\pm$ 2  & $-$290.0 & (1), (2), (9)  \\
    Fornax    & 02:39:59.3 & $-$34:26:57 & 2523 & 20   & 147 $\pm$ 12 & 838 $\pm$ 3  & $+$55.2  & (1), (4), (12) \\
    Leo I     & 10:08:28.1 & $+$12:18:23 & 175  & 5.5  & 254 $\pm$ 15 & 270 $\pm$ 2  & $+$282.9 & (1), (6), (13) \\
    Leo II    & 11:13:28.8 & $+$22:09:06 & 328  & 0.74 & 233 $\pm$ 14 & 171 $\pm$ 2  & $+$78.7  & (1), (7), (14) \\
    Sculptor  & 01:00:09.4 & $-$33:42:33 & 1360 & 2.3  & 86 $\pm$ 6   & 280 $\pm$ 1  & $+$111.4 & (1), (8), (12) \\
    Sextans   & 10:13:03.0 & $-$01:36:53 & 445  & 0.44 & 86 $\pm$ 4   & 413 $\pm$ 3  & $+$224.3 & (1), (5), (12) \\
    Ursa Minor& 15:08:08.5 & $+$67:13:21 & 318  & 0.29 & 76 $\pm$ 3   & 407 $\pm$ 2  & $-$246.9 & (1), (3), (10) \\
    
    \hline \hline
    \multicolumn{9}{l}{\footnotesize \textit{References:} (1) \citet{Munoz2018}; (2) \citet{Bonanos2004}; (3) \citet{Carrera2002}; } \\
    \multicolumn{9}{l}{\footnotesize (4) \citet{Pietrzynski2009}; (5) \citet{Lee2009}; (6) \citet{Bellazzini2004}; (7) \citet{Bellazzini2005};} \\
    \multicolumn{9}{l}{\footnotesize (8) \citet{Pietrzynski2008}; (9) \citet{Walker2015}; (10) \citet{Spencer2018}; (11) \citet{Fabrizio2016};} \\
    \multicolumn{9}{l}{\footnotesize (12) \citet{Walker2009a}; (13) \citet{Mateo2008}; (14) \citet{Koch2007}.} \\
  \end{tabular}
\end{table*}
\renewcommand{\arraystretch}{1}

Table \ref{Tab:observational_data} presents the observational properties of the eight dSphs analyzed in this study: Carina, Draco, Fornax, Leo I, Leo II, Sculptor, Sextans, and Ursa Minor. The table includes columns for the galaxy name, central photometric coordinates (R.A and decl.), the number of stars in the kinematic sample, heliocentric distance, projected half-light radius, systemic velocities, and associated references. For the purposes of this analysis, the projected half-light radius and systemic velocities are treated as fixed input parameters.

The stellar kinematic samples employed in this study are drawn from the literature. For Carina, Draco, Ursa Minor, Leo I, and Leo II dSphs, we adopt the spectroscopic datasets presented by \citet{Fabrizio2016, Walker2015, Spencer2018, Mateo2008}, and \citet{Spencer2017}, respectively. The kinematic data for Sextans, Sculptor, and Fornax are taken from \citet{Walker2009a, Walker2009b}. For each system, we retain only stars classified as members in the original analyses, and we do not attempt to redefine or reevaluate the membership selection. The membership criteria, therefore, follow those adopted in the respective observational studies.

We assume that stars behave as tracer particles within a gravitational potential dominated by dark matter, as dSphs are dark-matter dominated even in their inner regions \citep{McConnachie2012, BattagliaNipoti2022}. We further assume that the contribution of binary stars to the observed kinematics is minimal in dSphs\footnote{Despite being more dark matter dominated than dSphs, the low velocity dispersion in UFDs makes them more susceptible to contamination from unidentified binary stars \citep{McConnachie2010, Kirby2013, Kirby2017, Pianta2022, Chema2026}. Therefore, given the additional complexities in analyzing UFDs, this study is restricted to dSphs.} \citep{Minor2013, Spencer2017, Chema2023, Wang2023}.

Given the available kinematic samples for eight Milky Way dSphs, we fit the model parameters using the following likelihood function
\begin{equation}
    \log(\mathcal{L}_{\textrm{total}}) = \log(\mathcal{L}_{v_{\textrm{los}}}) + \log(\mathcal{L}_{c_{200}}),
    \label{eqn:totallikelihood}
\end{equation}
where the first term quantifies the likelihood of the observed stellar projected positions $R$ and LOS velocities $v_{\textrm{los}}$,
\begin{equation}
    \mathcal{L}_{v_{\rm los}} = \prod_{i=1}^{N} \frac{1}{ (\delta v_{\mathrm{los,}i}^2 + \sigma_{{\rm los,}i}^2)^{1/2}} f_s(w_i),
\end{equation}
with
\begin{equation}
w_i^2 = \frac{(v_{{\rm los,}i} - v_{\rm sys})^2}{\delta v_{\mathrm{los,}i}^2 + \sigma_{{\rm los,}i}^2}.
\end{equation}
The choice of the kernel function $f_s(w)$ is determined by the predicted LOS kurtosis $\kappa_{\rm los}$. Specifically, a uniform kernel is adopted when $\kappa_{\rm los} < 3$, while a Laplacian kernel is used when $\kappa_{\rm los} > 3$.

Our likelihood is constructed directly from the individual stellar velocity measurements rather than from binned velocity dispersion profiles. This unbinned approach preserves the full information content of the kinematic data and avoids the information loss that can arise when velocities are grouped into radial bins.

The second term in Equation (\ref{eqn:totallikelihood}) incorporates a prior on the halo concentration-–mass relation. This relation was originally reported in the CDM simulations \citep{Prada2012, Ishiyama2021}. In FDM cosmologies, the suppression of small-scale power delays halo formation, leading to deviations from the CDM concentration–-mass relation, as shown analytically by \citet{Laroche2022, Kawai2024} and demonstrated in simulations by \citet{Liao2025}. However, in the present analysis, the difference between the FDM and CDM concentration–-mass relations has a negligible impact on $m_\psi$ (see Appendix \ref{sec:recoveryposterior} for a demonstration). We therefore adopt, for simplicity, the empirical CDM concentration–mass relation for subhalos proposed by \citet{Moline2017},
\begin{equation}
\begin{split}
C_{200}(M_{200}, x_{\rm sub}) &= c_0 \\
& \times \left\{ 1 + \sum_{i=1}^3 \left[a_i \log_{10}\!\left(\frac{M_{200}}{10^8 h^{-1} ~ \mathrm{M_\odot}} \right) \right]^i \right\} \\
& \times \left[ 1 + b \log_{10}(x_{\rm sub}) \right],
\end{split}
\label{eqn:cmr}
\end{equation}
where the constants are $c_0 = 19.9$, $b = -0.54$, and $a_{i=1,2,3} = (-0.195, 0.089, 0.089)$. The dimensionless subhalo position parameter is defined as $x_{\rm sub} \equiv r_{\rm sub}/r_{\rm 200,host}$, where $r_{\rm sub}$ denotes the subhalo’s distance from the host halo center and $r_{\rm 200,host}$ is the virial radius of the host halo. We require the outer NFW halo to satisfy this concentration–-mass relation, yielding the log-likelihood
\begin{equation}
    -2 \log (\mathcal{L}_{c_{200}}) = \frac{[\log_{10}(c_{200}) - \log_{10}(C_{200})]^2}{\sigma^2_{c_{200}}},
\end{equation}
where $C_{200}$ is the median subhalo concentration given by Equation (\ref{eqn:cmr}), $\sigma_{c_{200}} = 0.13$, and $c_{200}$ is computed from the model parameters $(r_s, M_{200})$.
 
The full model contains five parameters over which we marginalize $\beta$, $m_\psi$, $M_{200}$, $\epsilon$, and $r_s$. We adopt log-flat priors for all parameters, with ranges
 \begin{enumerate}
     \item $-1 < -\log_{10}(1-\beta) < 1$,
     \item $-24 \leq \log_{10}(m_\psi/\textrm{eV}) \leq -18$,
     \item $6 \leq \log_{10}(M_{200}/\textrm{M}_\odot) \leq 12$,
     \item $-7 \leq \log_{10}(\epsilon) < 0$, which corresponds to $0 < r_t/r_c \leq 8.45$,
     \item $-1 \leq \log_{10}(r_s/r_h) \leq 3$.
 \end{enumerate}
The posterior distributions of the free parameters are sampled using the Markov Chain Monte Carlo (MCMC) method within the Metropolis–-Hastings framework \citep{Metropolis1953, Hastings1970}. We employ a custom-built code developed specifically for this analysis.

For each galaxy, the MCMC setup consists of 500 chains. 
We assess convergence using the Gelman--Rubin statistic $\hat{R}$ \citep{Gelman1992}, requiring $\hat{R} \lesssim 1.01$ for well-converged chains (see Appendix \ref{sec:GelmanRubin} for details).
However, this convergence diagnostic cannot be directly applied to our full analysis since each chain follows its own CHR and thus samples a different likelihood surface.
Therefore, we evaluated $\hat{R}$ using pilot MCMC runs assuming a single fixed CHR, for which multiple independent chains sample the same target posterior.
We employed 30 independent chains and selected representative CHRs with the largest and smallest values of the slope parameter $\eta$ for each run.
The chain length was then adjusted until all parameters satisfied $\hat{R} < 1.01$.
Based on these tests, we adopted a chain length of 20,000 samples for the main analysis, discarding the first 2000 samples as burn-in.
The final parameter constraints in the main analysis are derived by combining the posterior samples from all CHR realizations.

\section{Results} \label{Sec:Results}

\begin{figure}
    \centering
    \includegraphics[width=0.45\textwidth]{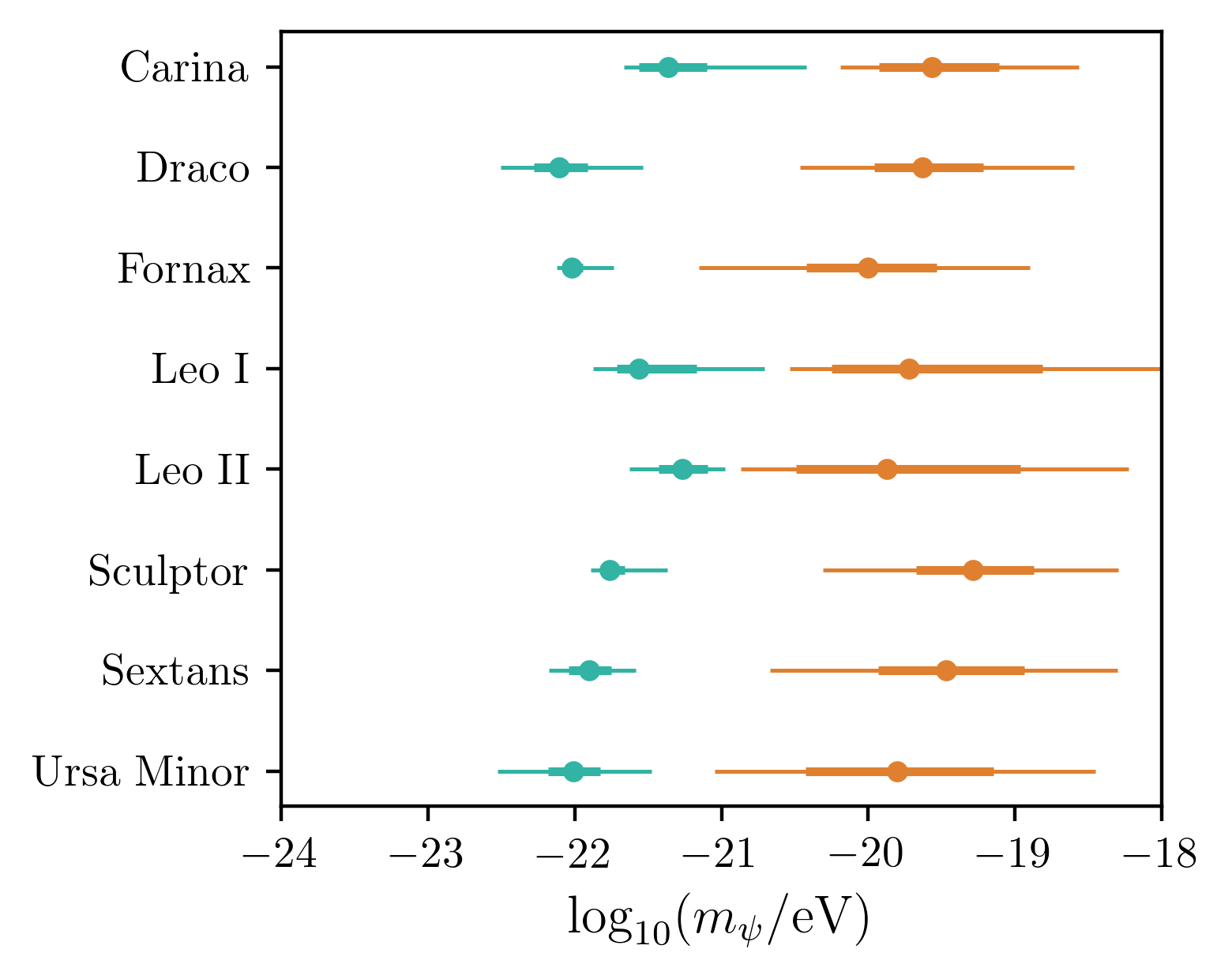}
    \caption{The estimated $m_\psi$ from this work for each galaxy. There are two high-probability regions, which correspond to two different DM core radii. Thick (thin) error bars indicate 68\% (95\%) credible interval.}
    \label{fig:mpsi}
\end{figure}

\renewcommand{\arraystretch}{1.1}
\begin{table*}
\centering
\setlength{\tabcolsep}{4.8pt}
\caption{Constraints on Model Parameters from Milky Way dSph Satellites, with uncertainties quoted at the 68\% credible interval.}
\label{tab:parameterconstraints}

\begin{tabular}{lccccccc}
\hline
\hline
Galaxy 
& $-\log_{10}(1-\beta)$
& $\log_{10}(m_\psi/\textrm{eV})$
& $\log_{10}(M_{200}/\textrm{M}_{\odot})$
& $r_t/r_c$
& $\log_{10}(r_s/\mathrm{kpc})$
& $\log_{10}(r_c/\mathrm{kpc})$
& $\mathrm{max}[\log(\mathcal{L_{\mathrm{total}}})]$\\
\hline
\multicolumn{8}{c}{\textbf{low-$m_\psi$}} \\
\hline
Carina & $0.08_{-0.14}^{+0.13}$ & $-21.37_{-0.20}^{+0.27}$ & $8.43_{-0.28}^{+0.31}$ & $2.59_{-1.09}^{+1.02}$ & $-0.18_{-0.18}^{+0.18}$ & $-0.51_{-0.20}^{+0.15}$ & $-3946.08$ \\
Draco & $-0.10_{-0.08}^{+0.09}$ & $-22.11_{-0.17}^{+0.20}$ & $9.49_{-0.29}^{+0.38}$ & $2.35_{-1.15}^{+1.30}$ & $0.07_{-0.14}^{+0.15}$ & $-0.14_{-0.12}^{+0.10}$ & $-1734.05$ \\
Fornax & $-0.03_{-0.05}^{+0.05}$ & $-22.02_{-0.05}^{+0.07}$ & $9.08_{-0.16}^{+0.26}$ & $2.79_{-0.93}^{+0.87}$ & $0.06_{-0.14}^{+0.13}$ & $-0.08_{-0.06}^{+0.04}$ & $-9810.03$ \\
Leo I & $-0.14_{-0.17}^{+0.19}$ & $-21.56_{-0.15}^{+0.39}$ & $8.73_{-0.27}^{+0.35}$ & $2.78_{-1.10}^{+0.93}$ & $-0.01_{-0.17}^{+0.16}$ & $-0.46_{-0.30}^{+0.11}$ & $-1195.50$ \\
Leo II & $0.39_{-0.25}^{+0.30}$ & $-21.27_{-0.16}^{+0.18}$ & $8.66_{-0.43}^{+0.72}$ & $2.91_{-0.93}^{+1.51}$ & $-0.04_{-0.22}^{+0.20}$ & $-0.68_{-0.14}^{+0.13}$ & $-599.18$ \\
Sculptor & $0.00_{-0.08}^{+0.07}$ & $-21.76_{-0.07}^{+0.10}$ & $8.96_{-0.23}^{+0.38}$ & $2.82_{-1.17}^{+0.78}$ & $-0.03_{-0.15}^{+0.15}$ & $-0.30_{-0.07}^{+0.05}$ & $-4984.06$ \\
Sextans & $0.06_{-0.14}^{+0.15}$ & $-21.90_{-0.14}^{+0.15}$ & $9.01_{-0.31}^{+0.48}$ & $2.95_{-1.27}^{+1.03}$ & $-0.03_{-0.16}^{+0.15}$ & $-0.09_{-0.12}^{+0.10}$ & $-1562.62$ \\
Ursa Minor & $0.28_{-0.20}^{+0.25}$ & $-22.01_{-0.17}^{+0.18}$ & $9.19_{-0.29}^{+0.47}$ & $2.74_{-1.21}^{+1.21}$ & $0.00_{-0.14}^{+0.15}$ & $-0.09_{-0.13}^{+0.13}$ & $-1166.34$ \\

\hline
\multicolumn{8}{c}{\textbf{high-$m_\psi$}} \\

\hline
Carina & $-0.16_{-0.12}^{+0.10}$ & $-19.56_{-0.36}^{+0.46}$ & $8.72_{-0.43}^{+0.43}$ & $3.52_{-0.74}^{+0.79}$ & $0.03_{-0.15}^{+0.15}$ & $-2.35_{-0.50}^{+0.46}$ & $-3948.55$ \\
Draco & $-0.15_{-0.09}^{+0.09}$ & $-19.63_{-0.33}^{+0.41}$ & $8.94_{-0.59}^{+0.56}$ & $3.62_{-0.71}^{+0.82}$ & $0.27_{-0.12}^{+0.12}$ & $-2.43_{-0.48}^{+0.43}$ & $-1737.18$ \\
Fornax & $-0.16_{-0.05}^{+0.04}$ & $-20.00_{-0.42}^{+0.47}$ & $8.83_{-0.50}^{+0.45}$ & $3.29_{-0.94}^{+0.70}$ & $0.24_{-0.10}^{+0.10}$ & $-2.03_{-0.49}^{+0.56}$ & $-9813.19$ \\
Leo I & $-0.32_{-0.12}^{+0.13}$ & $-19.72_{-0.53}^{+0.91}$ & $8.57_{-0.53}^{+0.65}$ & $3.15_{-0.99}^{+1.11}$ & $-0.06_{-0.21}^{+0.19}$ & $-2.25_{-0.84}^{+0.63}$ & $-1195.31$ \\
Leo II & $0.20_{-0.25}^{+0.30}$ & $-19.87_{-0.62}^{+0.91}$ & $7.96_{-0.68}^{+0.96}$ & $2.91_{-1.07}^{+1.34}$ & $-0.32_{-0.30}^{+0.34}$ & $-2.11_{-0.82}^{+0.74}$ & $-599.00$ \\
Sculptor & $-0.15_{-0.05}^{+0.05}$ & $-19.29_{-0.39}^{+0.42}$ & $8.76_{-0.53}^{+0.41}$ & $3.64_{-0.81}^{+0.67}$ & $0.17_{-0.12}^{+0.11}$ & $-2.63_{-0.43}^{+0.50}$ & $-4982.02$ \\
Sextans & $-0.13_{-0.10}^{+0.11}$ & $-19.47_{-0.46}^{+0.53}$ & $8.63_{-0.32}^{+0.37}$ & $3.41_{-0.75}^{+0.67}$ & $0.03_{-0.13}^{+0.14}$ & $-2.36_{-0.57}^{+0.51}$ & $-1566.29$ \\
Ursa Minor & $0.09_{-0.13}^{+0.20}$ & $-19.80_{-0.62}^{+0.65}$ & $8.89_{-0.55}^{+0.55}$ & $3.37_{-1.25}^{+1.12}$ & $0.02_{-0.17}^{+0.15}$ & $-2.21_{-0.70}^{+0.77}$ & $-1167.47$ \\

\hline
\end{tabular}
\end{table*}
\renewcommand{\arraystretch}{1.1}

Figure \ref{fig:mpsi} presents the inferred $m_\psi$ for each analyzed dSph, with thick and thin error bars indicating the 68\% and 95\% credible intervals, respectively. All galaxies exhibit a bimodal posterior distribution in $m_\psi$, with systems that have larger kinematic samples tending to show more clearly separated high-probability regions. In the figure, the solution peaking at lower $m_\psi$ is shown in green, while the solution favoring higher $m_\psi$ is shown in orange.\footnote{The share of posterior probability mass contained in the two regimes ranges from low-$m_\psi$/high-$m_\psi$ = 0.23/0.77 to 0.91/0.09.} We summarize the constraints on the model parameters obtained in this analysis in Table \ref{tab:parameterconstraints}, where the quoted uncertainties correspond to the 68\% credible intervals.\footnote{Throughout this work, we report constraints derived from a Bayesian analysis. The quoted intervals correspond to credible regions for $m_\psi$ inferred from the posterior distribution within the adopted FDM halo model, and should therefore be interpreted as parameter constraints within this framework rather than as exclusion limits obtained under different statistical interpretations.} The full posterior distributions of all parameters and recovery of the observed kinematics are presented in Appendix \ref{sec:recoveryposterior}.

The bimodality in $m_\psi$ arises primarily because the steep density decline of the soliton profile, $\rho(r) \propto r^{-16}$ between $r_c$ and $r_t$, is not favored by the stellar kinematic data. Such a rapidly declining density profile produces significantly lower LOS velocity dispersions than both the NFW profile and the inner core of the soliton profile (see Appendix~\ref{sec:recoveryposterior} for a demonstration). Consequently, the model tends to avoid configurations in which many kinematic samples lie between $r_c$ and $r_t$. This behavior gives rise to two viable configurations: either (i) most kinematic samples reside within the soliton core, or (ii) most kinematic samples lie in the NFW-like region.

\begin{figure}
    \centering
    \includegraphics[width=0.4\textwidth]{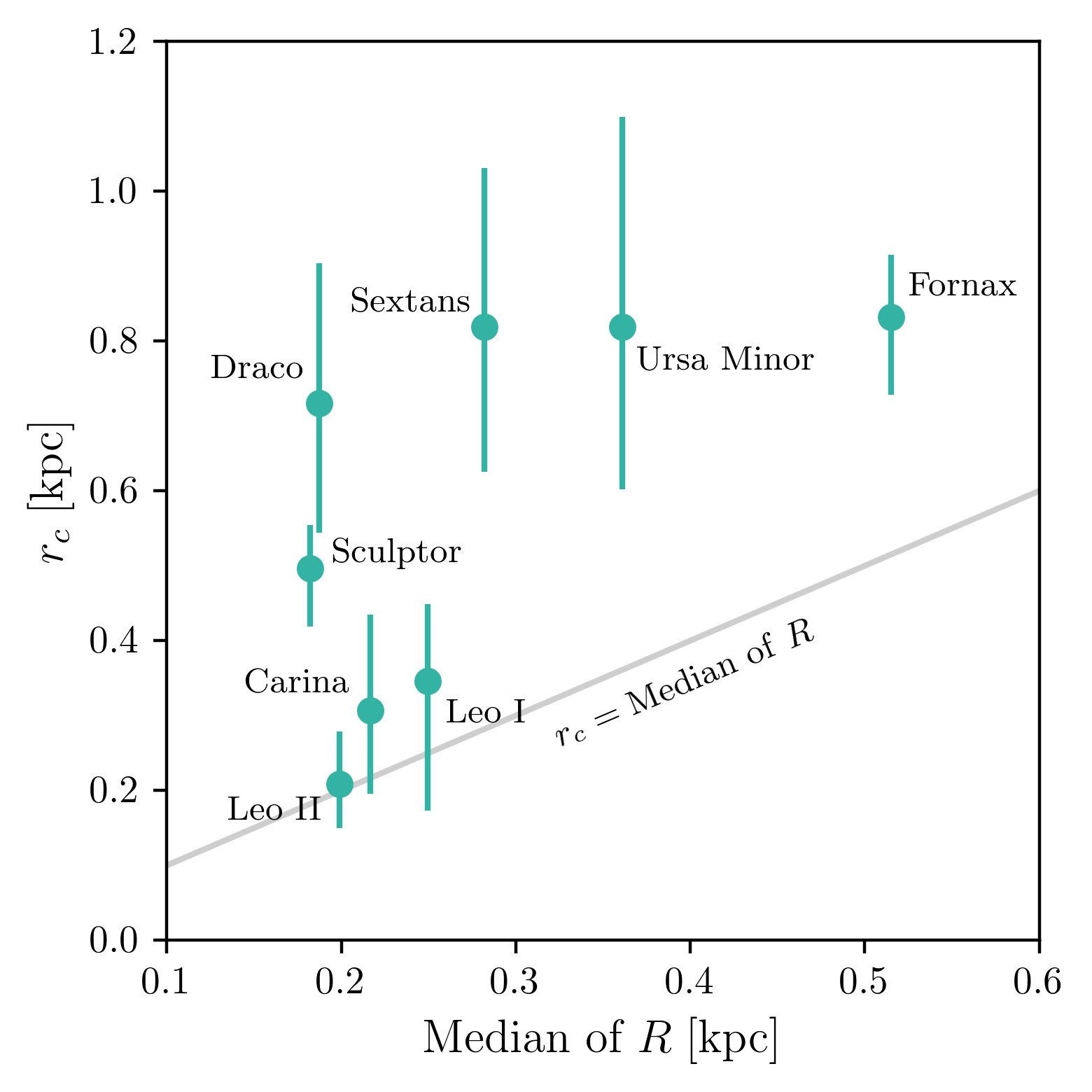}
    \caption{Core radius $r_c$ of the low-$m_\psi$ solutions vs. the median projected galactocentric distance of the kinematic samples. Error bars denote the 68\% credible intervals. The solid gray line represents $r_c = \mathrm{median}(R)$ to guide the readers. For the low-$m_\psi$ solution, the dark matter core radius must be sufficiently large to encompass most of the kinematic tracers within the core region.}
    \label{fig:rc_vs_Rmedian}
\end{figure}

The first scenario requires a substantially large DM core, giving rise to the posterior peak at $m_\psi \sim 10^{-22} ~ \mathrm{eV}$. This requirement is reflected in Figure \ref{fig:rc_vs_Rmedian}, which compares the inferred $r_c$ for the low-$m_\psi$ solution with the median projected radius of the stellar kinematic samples. The diagonal gray line indicates $r_c=\mathrm{median}(R)$ for reference. For all galaxies in our sample, $r_c$ exceeds the median stellar radius, thereby placing the majority of kinematic tracers within the soliton core. These results highlight the importance of identifying kinematic samples that extend well beyond the currently available radial coverage, such as those that will be provided by the Subaru Prime Focus Spectrograph (PFS) survey \citep{Takada2014, Tamura2016, Chiba2026}, DESI \citep{Cooper2023}, WEAVE \citep{Dalton2012, Jin2024}, and 4MOST \citep{Skuladottir2023}. If stellar kinematics at larger radii are inconsistent with a sharp decline in $\rho(r)$, the model requires an even larger $r_c$ to reproduce the observations.

In contrast, the second configuration corresponds to the high-probability region at relatively high particle masses, $m_\psi \sim 10^{-20} ~ \mathrm{eV}$.
In this case, the inferred transition radius lies between 10 and 30 pc, which does not exceed the radius enclosing 1\% of the kinematic samples in each galaxy.
The appearance of a high-probability region in this $m_\psi$ range highlights the role of an NFW-like envelope in shaping the kinematic inference of FDM halos \citep{HayashiObata2020}.
In its absence, the model preferentially admits only the low-$m_\psi$ solution, effectively restricting the favored parameter range.
A modeling assumption in which halos are described solely by a soliton profile has been adopted in earlier studies such as \citet{Chen2017}, \citet{HayashiObata2020}, and \citet{Martino2023}, providing a useful point of comparison for understanding how the inclusion of an outer NFW envelope affects the inferred constraints.

The bimodality in $m_\psi$ is not unique to this work and has been reported in previous studies. For example, \citet{HayashiObata2020} analyzed the same dSph galaxies considered here using a nonspherical dynamical model and found that a bimodal posterior in $m_\psi$ emerges when the halo model is extended from a pure soliton profile to a soliton plus NFW profile.
Similarly, \citet{Goldstein2022} reported bimodality in all six dSphs included in their analysis, all of which are also part of our sample.
Although both studies adopted a one-to-one CHR, the bimodality likely shares a common physical origin with that found here.
The inclusion of CHR diversity does not create this bimodality, but it does influence the relative prominence and breadth of the two modes.
Insight into this behavior can be gained from the distribution of MCMC samples in the parameter space of the CHR and halo properties.

\begin{figure}
    \centering
    \includegraphics[width=0.49\textwidth]{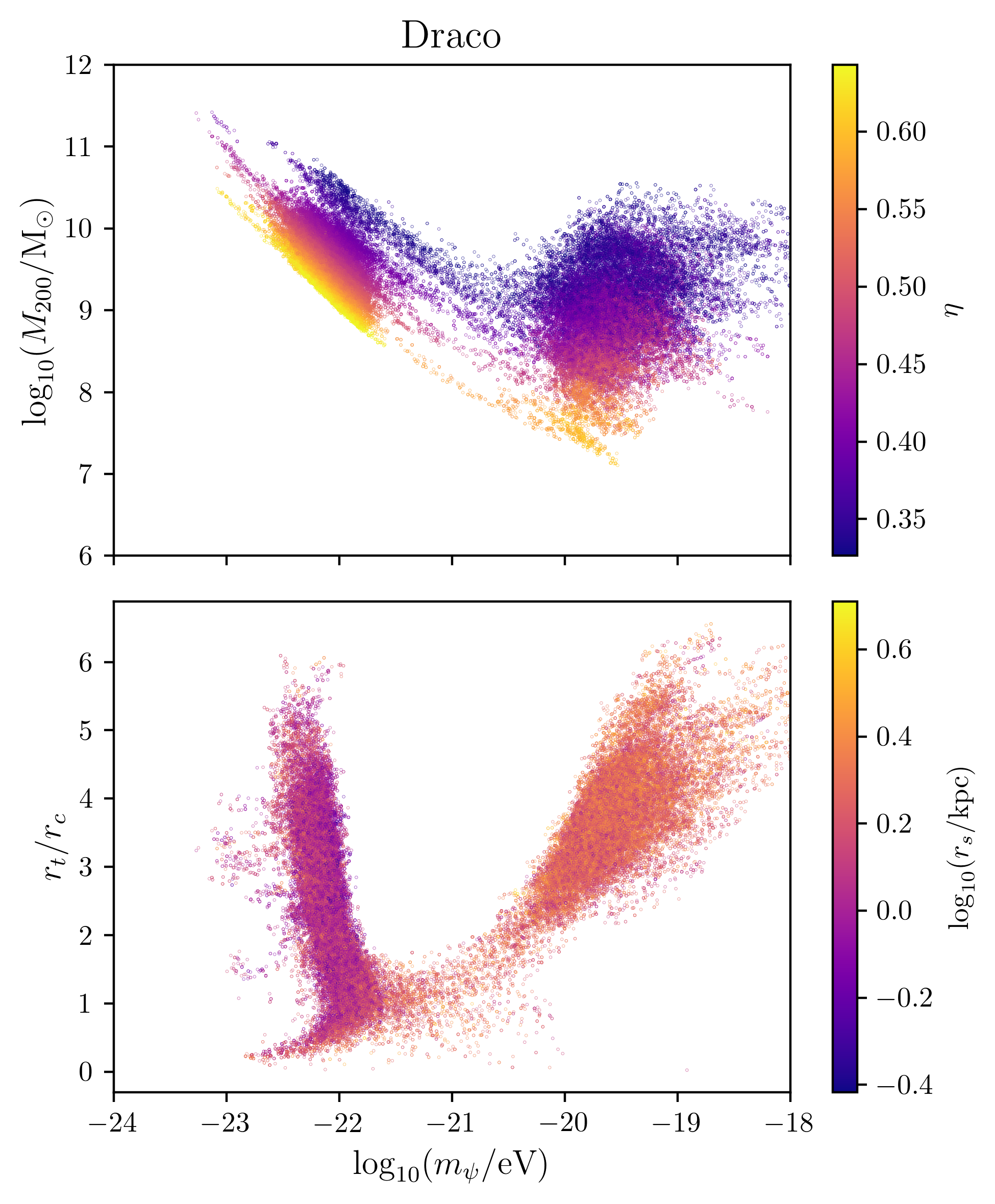}
    \caption{Upper panel: MCMC samples of the estimated $\log_{10}(M_{200}/\mathrm{M_\odot})$ vs. $\log_{10}(m_\psi/\textrm{eV})$ for Draco as representative. The colors are coded based on the slope of the CHR $\eta$. Lower panel: MCMC samples of the estimated $r_t/r_c$ vs. $\log_{10}(m_\psi/\textrm{eV})$ with the colors coded by $\log_{10}(r_s/\textrm{kpc})$.}
    \label{fig:mpsi-M200_B_heatmap}
\end{figure}

To illustrate the impact of CHR diversity, Figure~\ref{fig:mpsi-M200_B_heatmap} shows the distribution of MCMC samples for Draco, which serves as a representative example.
The remaining galaxies show broadly similar behavior.
The upper panel presents the sample distribution in the $\log_{10}(M_{200}/\mathrm{M_\odot})$ versus $\log_{10}(m_\psi/\textrm{eV})$ plane, with colors indicating the CHR slope, $\eta$.
Notably, the slope reported by \citet{Schive2014b} lies close to the lower edge of the distribution, corresponding to $\eta \approx 0.33$.
A degeneracy between $m_\psi$ and $M_{200}$, previously highlighted by \citet{Goldstein2022} in connection with the bimodality, is visible as diagonal patches extending from the upper left to the lower right.
This degeneracy is expected and intrinsic to the CHR.
However, it weakens at higher $m_\psi$, where the soliton core mass constitutes a smaller fraction of the total halo mass, as implied by Equation (\ref{eqn:core_halo_mass_relation_jowett}).

The color coding in the upper panel of Figure \ref{fig:mpsi-M200_B_heatmap} further illustrates that, in the absence of diversity in the CHR, these diagonal structures would appear as a single narrow strip (see, for example, the posterior distribution reported by \citet{Chen2017}).
It is important to note that solutions with smaller $\eta$ are suppressed in the low-$m_\psi$ regime but dominate at higher $m_\psi$, and vice versa.
This behavior can be understood from Equations~(\ref{eqn:Mc-rc_relation}) and (\ref{eqn:core_halo_mass_relation_jowett}), which imply that the central density scales approximately as $\rho_c \propto m_\psi^{6\eta}$.
Therefore, a small $\eta$ can mitigate the rapid increase in $\rho_c$, leading CHRs with small $\eta$ to dominate the high-$m_\psi$ solutions.
Even so, for $\log_{10}(m_\psi/\textrm{eV}) \gtrsim -19$, $\rho_c$ becomes excessively large, such that only a small number of MCMC samples remain consistent with the data.
In contrast, at low $m_\psi$, a small $\eta$ tends to produce core densities that are too low to reproduce the observed stellar kinematics.
This explains why CHRs with larger $\eta$ both dominate and broaden the low-$m_\psi$ mode.


\begin{figure*}
  \centering
  \includegraphics[width=0.9\textwidth]{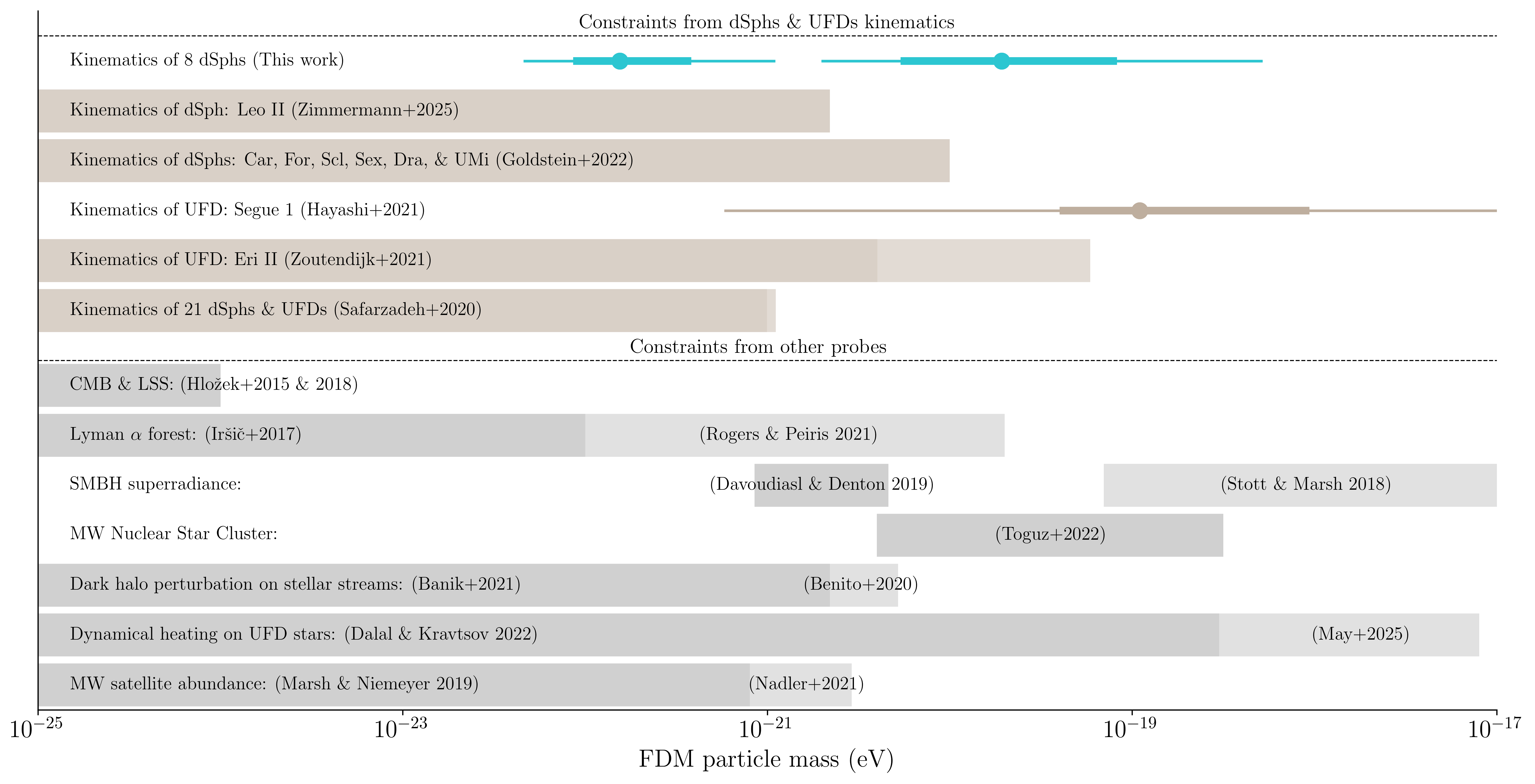}
  \caption{Constraints on the fuzzy dark matter particle mass. Points with error bars indicate the ranges of $m_\psi$ consistent with the data within the 68\% (thick error bars) and 95\% (thin error bars) credible intervals. Shaded bands denote excluded ranges of $m_\psi$. The upper section, in cyan and brown, shows constraints derived from this work and from other kinematic analyses of dwarf galaxies, respectively. Darker and lighter shades on the bands show exclusion at 95\% and 68\% credible intervals, respectively.  The lower part, in gray, summarizes constraints obtained from a variety of independent astrophysical and cosmological probes as indicated. Different shades of gray in the same row correspond to different works.}
  \label{fig:fdm_mass_constraints}
\end{figure*}

In the lower panel of Figure~\ref{fig:mpsi-M200_B_heatmap}, we show the MCMC samples in the $r_t/r_c$ versus $\log_{10}(m_\psi/\textrm{eV})$ plane to examine the behavior of $r_t$ in the two $m_\psi$ modes.
The colors indicate $\log_{10}(r_s/\textrm{kpc})$.
In the low-$m_\psi$ mode, no clear correlation is apparent.
This is likely because the transition radius is located at relatively large radii, $\gtrsim r_h$, where only a small fraction of the kinematic samples probe the gravitational potential.
In contrast, a correlation emerges in the high-$m_\psi$ mode, with larger $m_\psi$ corresponding to larger values of $r_t/r_c$.
This trend can be understood as follows.
At $\log_{10}(m_\psi/\textrm{eV}) \approx -21$, the DM core radius is $\sim 0.1$ kpc, and large $r_t/r_c$ are disfavored because an extended region with $\rho(r) \propto r^{-16}$ would encompass a substantial fraction of the kinematic samples.
As $m_\psi$ increases, the core radius decreases as $r_c \propto m_\psi^{-(3\eta + 1)/2}$, while the central density increases as $\rho_c \propto m_\psi^{6\eta}$ (with $\eta \approx 0.3$--$0.6$).
The steep density decline between $r_c$ and $r_t$ then helps mitigate the increasingly dense central core, making larger values of $r_t/r_c$ more favorable.
This trend persists provided that $r_t$ remains smaller than the projected radii of some of the innermost kinematic samples.

Since the posterior distribution of $m_\psi$ is bimodal, it is important to check whether the two modes are robust features of the inference or whether they are sensitive to prior assumptions. We therefore performed two complementary checks addressing different aspects of this question. First, we tested the sensitivity to the adopted prior range by repeating the Leo II analysis with a wider log-flat prior, $-25 < \log_{10}(m_\psi/{\rm eV}) < -16$, finding that the locations of both modes remain stable, shifting by less than $0.5\%$. This indicates that the mode locations are not simply fixed by the boundaries of the adopted $m_\psi$ prior. Second, we compared the maximum likelihood values of the two branches, as reported in Table~\ref{tab:parameterconstraints}, to assess whether both modes provide acceptable fits to the data. The differences, $|\Delta \log \mathcal{L}_{\rm total}^{\max}| = 0.18$--$3.67$, remain moderate across the sample, indicating that both branches provide comparably good fits to the data. These checks do not constitute a complete diagnostic of prior-volume effects, which would require a dedicated analysis such as profile likelihoods. However, they show no clear indication that the bimodality is caused by the prior boundaries or that one of the two modes is merely a poor-fit posterior tail.

\section{Discussion} \label{Sec:Discussion}

We summarize our results in Figure~\ref{fig:fdm_mass_constraints}, which also compiles current constraints on the FDM particle mass. Points with error bars indicate the ranges of $m_\psi$ consistent with the data within the 68\% (thick error bars) and 95\% (thin error bars) credible intervals, while shaded bands indicate excluded regions. Constraints from this work, shown in cyan, are derived by directly combining the posterior distributions of $m_\psi$ from all eight dSphs, thereby preserving a conservative statistical treatment throughout the analysis.
We find two statistically allowed ranges of $m_\psi$ from the observed kinematics: $\log_{10}(m_\psi/\mathrm{eV}) = -19.72^{+0.64}_{-0.56}$ and a narrower low-mass window at $\log_{10}(m_\psi/\mathrm{eV}) = -21.81^{+0.39}_{-0.26}$, each corresponding to the 68\% credible interval.



Other bounds derived from kinematical analyses of dwarf galaxies \citep{Safarzadeh2020, Hayashi2021, Zoutendijk2021, Goldstein2022, Zimmermann2025} are shown in brown. The darker and lighter shades indicate exclusion at the 95\% and 68\% credible intervals, respectively. The lower portion of the figure includes constraints from a variety of astrophysical and cosmological probes, including the cosmic microwave background and large-scale structure \citep{Hlozek2015, Hlozek2018}, the Lyman-$\alpha$ forest \citep{Irsic2017, RogersPeiris2021}, supermassive black hole superradiance \citep{Stott2018, Davoudiasl2019}, kinematics of the Milky Way nuclear star cluster \citep{Toguz2022}, perturbations of stellar streams by dark halos \citep{Benito2020, Banik2021}, dynamical heating of UFD stars \citep{Dalal2022, May2025}, and Milky Way satellite abundances \citep{Marsh2019, Nadler2021}. In this part, different tones of gray refer to different works.

Compared with previous constraints derived from dSph and UFD kinematics, the high-$m_\psi$ solution obtained in this work is broadly consistent with earlier results. The low-$m_\psi$ solution, on the other hand, remains statistically allowed in our analysis due to the inclusion of diversity in the CHR, as illustrated in Figure~\ref{fig:mpsi-M200_B_heatmap}. However, if the parameter inference ultimately favored this region of parameter space, it would be more challenging to reconcile with several previous kinematic bounds. 

To understand the origin of this difference, it is instructive to compare our analysis with previous studies that adopted similar dynamical modeling approaches. A particularly relevant example is \citet{Hayashi2021}, who employed a comparable framework to analyze the kinematics of UFD galaxies. Their modeling assumed a one-to-one CHR based on \citet{Schive2014b} and relied exclusively on second-order velocity moments. In the present analysis, we extend this framework by allowing for the CHR diversity suggested by simulations and by incorporating fourth-order LOS velocity moments. These modifications lead to a qualitatively different posterior structure, most notably by altering the relative prominence and breadth of the two $m_\psi$ modes. This comparison illustrates that relatively modest changes in the modeling assumptions can have a significant impact on the inferred constraints on the FDM particle mass.

Another common feature of our analysis and that of \citet{Hayashi2021} is the use of unbinned stellar velocity measurements to construct the likelihood, rather than velocity-moment profiles derived from radial binning. By preserving the full information content of the kinematic data, this approach may improve sensitivity to smaller soliton cores. Indeed, despite the comparatively small kinematic samples available for UFDs, \citet{Hayashi2021} obtained a stringent constraint from Segue I, finding a favored mass range of $m_\psi = 1.1^{+8.3}_{-0.7} \times 10^{-19}~\mathrm{eV}$ within 68\% credible interval. This value lies at the upper end of the high-$m_\psi$ solutions inferred in our analysis.

Potential systematic discrepancies between constraints derived from dSphs and UFDs have been discussed previously. For example, \citet{Safarzadeh2020} combined arguments based on the CHR and dynamical friction, concluding that no single range of $m_\psi$ can simultaneously reproduce dSph kinematics while avoiding implausibly massive UFD host halos ($M \sim 10^{11}$--$10^{12}\,\mathrm{M_\odot}$). Our results suggest that allowing the intrinsic diversity in the CHR can shift the lower bound toward lower $m_\psi$, because larger $\eta$ than those proposed by \citet{Schive2014b} permit lower halo masses without requiring large $r_c$. Whether this shift can alleviate the differences between dSph and UFD constraints remains to be investigated in future work.

In a complementary line of work, \citet{Zimmermann2025} derive a constraint using stellar kinematics in Leo II dSph.
Their approach is conceptually distinct from standard Jeans modeling and does not rely on assumptions of dynamical equilibrium or the CHR.
The authors reconstruct the galaxy’s data-driven phase-space structure and compare it to ensembles of FDM wave function realizations consistent with Schrödinger--Poisson dynamics. Agreement is assessed using a maximum mean discrepancy test against the inferred stellar phase-space distribution. The authors find that $m_\psi < 2.2 \times 10^{-21}~\mathrm{eV}$ produces excess small-scale structure and is excluded at the 95\% credible interval, which is compatible with the result found here.

Despite the overall agreement with constraints from CMB, LSS, and supermassive black hole superradiance, it is informative to compare our results with bounds derived from other astrophysical probes. One example comes from constraints based on the abundance of Milky Way satellite galaxies \citep{Marsh2019, Nadler2021}. The high-$m_\psi$ solution obtained in this work remains fully compatible with these bounds. The low-$m_\psi$ solution, while still statistically allowed, lies closer to the lower edge of the parameter space favored by satellite-count analyses and is therefore somewhat more challenged by these constraints. In particular, $m_\psi$ values that generate dark matter cores large enough to explain dSph kinematics also suppress the formation of lower-mass halos, reducing the predicted number of satellites. As a result, low-$m_\psi$ solutions arise in a regime where the suppression of small-scale structure becomes increasingly important. This situation is reminiscent of the well-known “catch-22” discussed for simple warm dark matter models \citep{Maccio2012}, although in the present case, the allowed ranges still overlap and the comparison depends on the modeling assumptions entering the different analyses.

Another useful comparison comes from the analysis of \citet{Toguz2022}, who studied stellar kinematics in the nuclear star cluster surrounding the Milky Way’s central supermassive black hole. Taken at face value and within the assumptions of that analysis, their results favor $m_\psi$ outside the range corresponding to the high-$m_\psi$ solutions obtained here. In their excluded $m_\psi$ range, the soliton core is expected to leave a detectable imprint on stellar motions, which has not been observed. For lower $m_\psi$, the soliton core density is too low to affect stellar dynamics, while for higher $m_\psi$, the soliton core mass becomes negligible compared to that of the supermassive black hole. We note, however, that the analysis of \citet{Toguz2022} assumes isotropic stellar orbits, $\beta = 0$. A more general dynamical model allowing for anisotropic velocity distributions may yield less restrictive constraints and would be required for a more robust comparison.

Some of the strongest constraints are reported by \citet{Dalal2022} and subsequently updated by \citet{May2025}, both of which examine the spatial distribution of stars in UFDs. These studies analyze the dynamical heating induced by stochastic FDM granules—interference patterns in the dark matter density field that can transfer energy to stars and progressively inflate their orbits. Based on this mechanism, \citet{Dalal2022} exclude $m_\psi < 3 \times 10^{-19} \, \mathrm{eV}$, while \citet{May2025}, who include the previously neglected nonlinear phenomenon in the simulations, derive a more stringent bound of $m_\psi < 8 \times 10^{-18} \, \mathrm{eV}$. However, the efficiency of granule-induced heating is not yet fully settled and continues to be explored, with uncertainties in both the theoretical modeling and its implementation in simulations (see, e.g., \citet{Eberhardt2025c} who studied the effect of stellar gravity, tidally stripped halos, and relative size of soliton to stellar distribution on this heating). Moreover, these constraints are inferred from long-term dynamical evolution rather than from instantaneous kinematical measurements, and may therefore be more sensitive to unmodeled processes in the evolutionary history of UFDs \citep{Zimmermann2025}.

It is also worth noting that constraints on FDM have been derived from kinematic studies of rotation-dominated galaxies \citep{Bar2018,Bernal2018,Chan2021,Bar2022,Banares2023,Khelashvili2023,Farisy2025}. These analyses generally model galaxy rotation curves by estimating the contributions of individual mass components, such as stars, gas, and dark matter. The inferred dark matter core properties are then used to constrain the allowed range of $m_\psi$. However, constraints derived from rotation-curve analyses can vary significantly across different galaxies. In some cases, preferred parameter values are not fully consistent across galaxy samples \citep{Khelashvili2023,Farisy2025}, while different analyses may exclude parameter regions that are favored in others \citep{Bar2022,Banares2023}. These differences may reflect the diversity of galaxy properties, as well as systematic uncertainties in mass modeling and the complex baryonic physics of disk galaxies.


Overall, the high-$m_\psi$ solution obtained in this work remains broadly compatible with most existing astrophysical bounds. The low-$m_\psi$ solution, while still statistically allowed, lies closer to the edge of several constraints discussed above and is therefore more strongly challenged by those probes. It is important to emphasize, however, that the presence of two allowed regions in our analysis reflects a statistical feature of the likelihood rather than two distinct physical models. Future analyses incorporating improved data or refined modeling may help determine whether one of these regions is ultimately favored. Another factor that may influence the inferred posterior is the particular set of dwarf galaxies included in the analysis. Individual systems can prefer different halo masses and therefore different regions of $m_\psi$ parameter space, such that the combined constraint may be driven by a subset of galaxies with especially strong constraining power. This behavior has been noted in previous kinematic analyses of dwarf galaxies, where different systems yield different preferred mass scales (see, e.g., \citet{Hayashi2021}). Future studies with larger samples and improved kinematic data will help clarify the extent to which such sample dependence affects the inferred constraints on $m_\psi$.

More generally, most existing bounds on $m_\psi$, including those presented in this work, rely on specific modeling assumptions and may therefore be affected by systematic uncertainties that are difficult to quantify. This applies not only to kinematic studies of dwarf galaxies but also to several other astrophysical probes discussed above. In each case, the inferred limits depend on assumptions about the structure and dynamical state of the systems being analyzed, as well as on the modeling of the dark matter distribution itself. As illustrated here through the inclusion of diversity in the CHR, different modeling choices can lead to significantly different inferred constraints. This sensitivity to analysis assumptions motivates the approach adopted in this work, where we explicitly explore how specific modeling ingredients affect the inferred limits on $m_\psi$. Systematically testing these assumptions is essential for assessing the robustness of current bounds and for understanding how results obtained with different probes should be compared.

In addition to these modeling choices within the dark-matter-only framework, baryonic processes may also play an important role. For example, supernova feedback has been shown to increase the size of dark matter cores and reduce their central densities \citep{Robles2024}. If baryonic feedback significantly affects the internal structure of dSphs, analyses that neglect these effects—-including the present one—-may underestimate the original central density, potentially shifting the inferred upper limits toward higher $m_\psi$.

We emphasize that our analysis relies on the assumptions of dynamical equilibrium and spherical symmetry. Although FDM halos are generally found to be more spherical than their CDM counterparts \citep{Schive2014a}, adopting a nonspherical framework would provide a more general and realistic description of the dynamics. Furthermore, simulations indicate that the soliton core undergoes stochastic motion \citep{Veltmaat2018, Schive2020, Chowdhury2021}, which may influence stellar kinematics but is not accounted for in this analysis.

\section{Conclusions} \label{Sec:Conclusions}

In this work, we investigate the impact of diversity in the CHR on the inference of the FDM particle mass from stellar kinematics in dSph galaxies. Our analysis extends previous studies by incorporating both the diversity of the CHR suggested by cosmological simulations and higher-order information in the form of the fourth moment (kurtosis) of the LOS velocity distribution. This additional dynamical information helps mitigate degeneracies present in standard Jeans analyses based solely on the velocity dispersion.

We model the dark matter halo as a soliton core embedded in an NFW halo and apply this framework to stellar kinematic data of eight Milky Way's dSphs: Carina, Draco, Fornax, Leo~I, Leo~II, Sculptor, Sextans, and Ursa Minor. By combining the posterior distributions inferred for each system, we find two preferred regions for the FDM particle mass at the 68\% credible interval: $\log_{10}(m_\psi/\mathrm{eV}) = -19.72^{+0.64}_{-0.56}$ and $\log_{10}(m_\psi/\mathrm{eV}) = -21.81^{+0.39}_{-0.26}$. These ranges reflect the bimodality of the posterior distribution obtained from the current data and analysis choices.

The high-$m_\psi$ solution is broadly compatible with most existing astrophysical constraints, while the low-$m_\psi$ region lies closer to the edge of several bounds discussed in the literature. Importantly, the presence of these two solutions reflects a statistical feature of the likelihood rather than two distinct physical models. Future analyses incorporating improved data and refined modeling may help determine whether one of these regions is ultimately favored.

More generally, our results highlight the sensitivity of inferred FDM constraints to modeling assumptions entering the dynamical analysis. Even relatively modest changes in the treatment of the CHR or the inclusion of higher-order velocity moments can alter the inferred posterior structure and the resulting constraints on $m_\psi$. This sensitivity is consistent with previous studies showing that different analysis choices, data selections, or modeling assumptions can lead to substantially different limits on the particle mass.

Systematically exploring these assumptions is therefore essential for assessing the robustness of current bounds on FDM. In particular, future observations with next-generation spectroscopic surveys such as the Subaru PFS will provide significantly improved stellar velocity measurements for dwarf galaxies. These data will enable more precise dynamical analyses and offer an opportunity to test the modeling assumptions underlying current constraints, helping to clarify the allowed parameter space of FDM.

\begin{acknowledgments}
We thank Yutaka Hirai for generously providing access to the computational resources used in this work. We are also grateful to Shunichi Horigome, Victor Robles, Hsi-Yu Schive and group members, Tzihong Chiueh, and Andrew Cooper for fruitful discussions. This work was supported by JSPS KAKENHI grant Nos. JP24K00669~(M.C., K.H.), JP25H00394, JP25K24687 (M.C.), JP25H01553, JP26H02044, JP26K07153~(K.H.).
\end{acknowledgments}

\begin{contribution}
All authors contributed equally to this work.
\end{contribution}

\software{\texttt{Matplotlib} \citep{Hunter2007},
          \texttt{NumPy} \citep{VanDerWalt2011},
          \texttt{SciPy} \citep{Virtanen2020}
          }

\appendix \label{sec:Appendix}

\section{Posterior Distributions and Recovery of kinematics} \label{sec:recoveryposterior}
\begin{figure*}
\centering
\gridline{
\fig{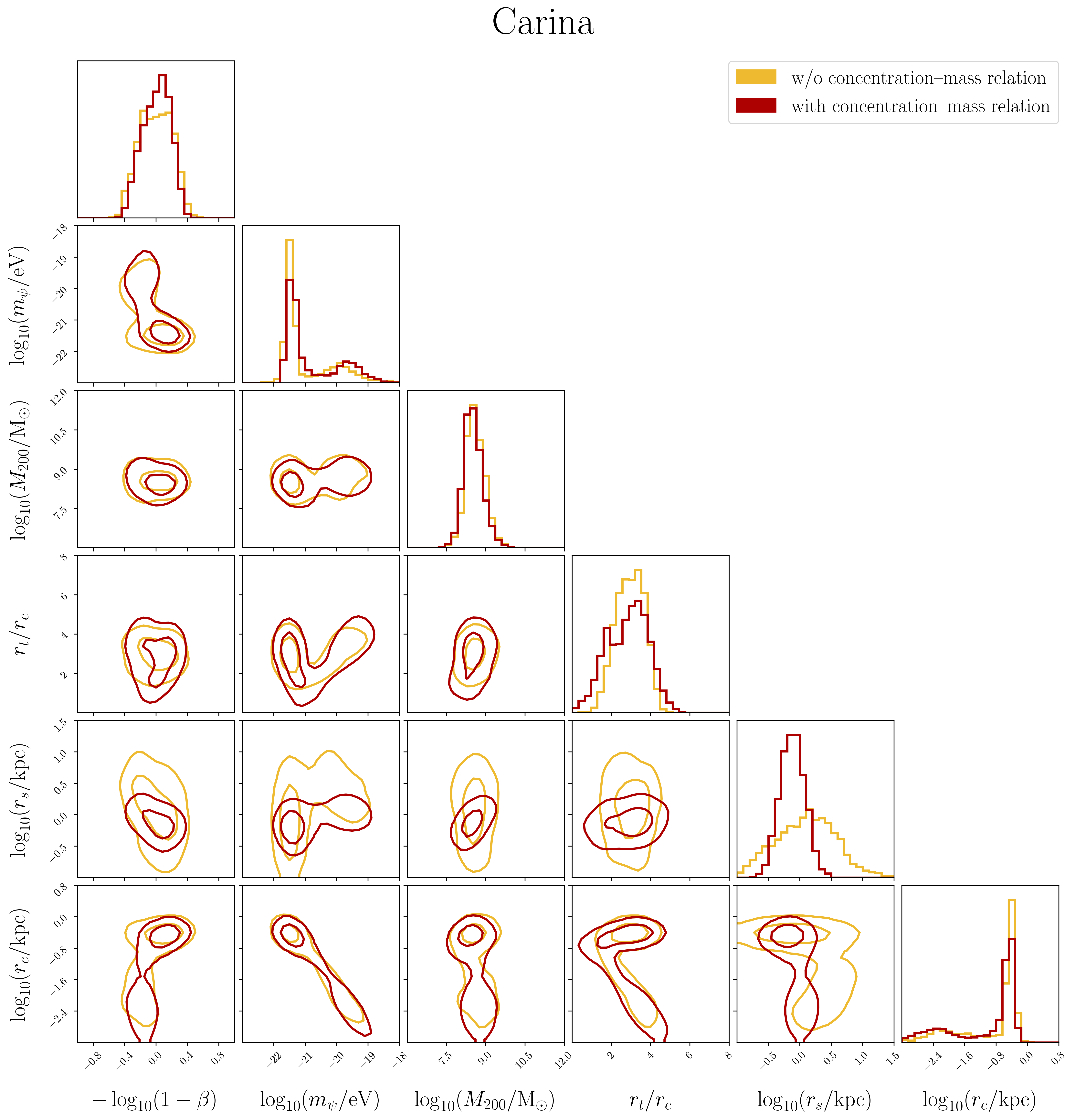}{0.45\textwidth}{}
\fig{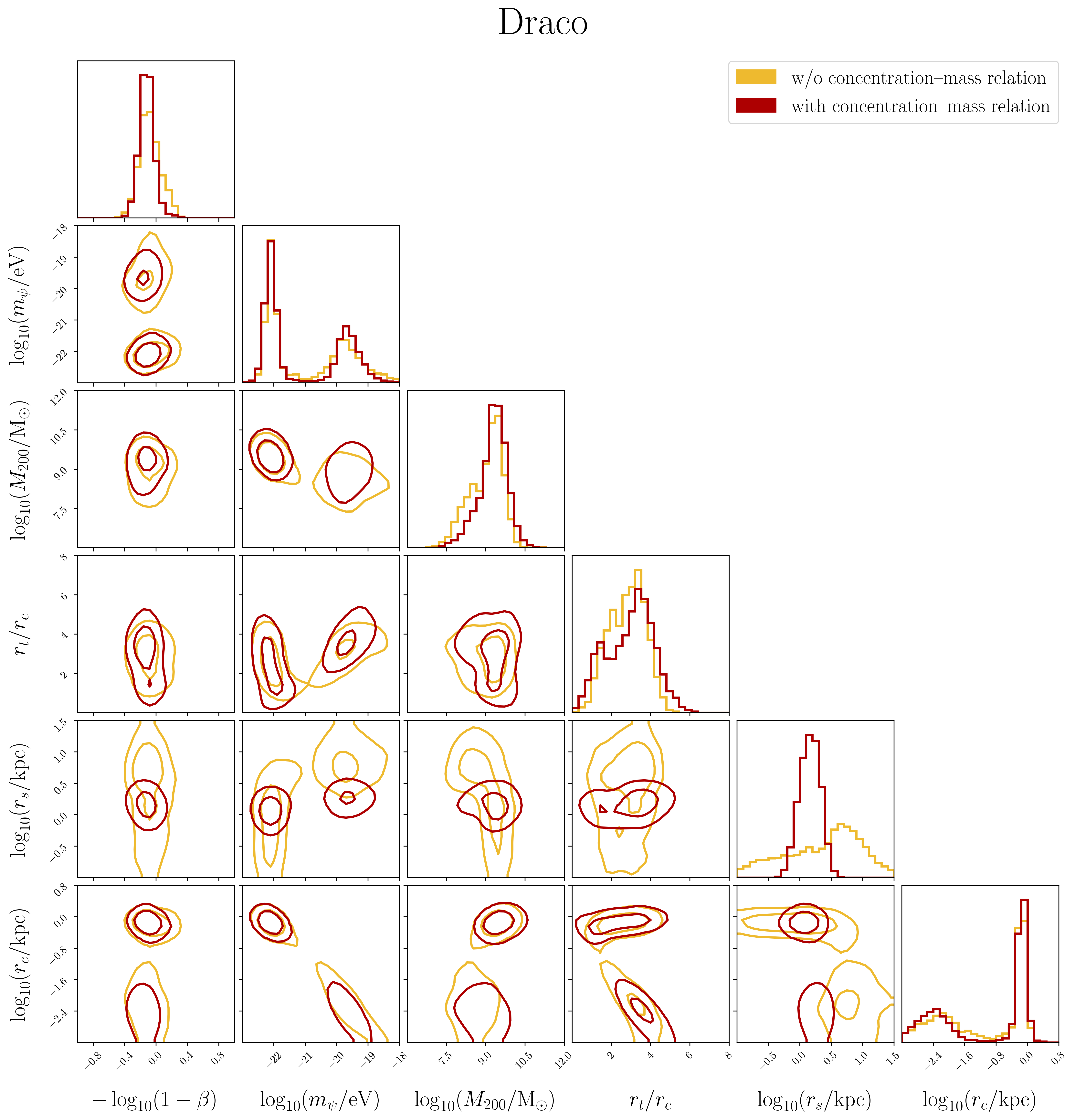}{0.45\textwidth}{}
}
\gridline{
\fig{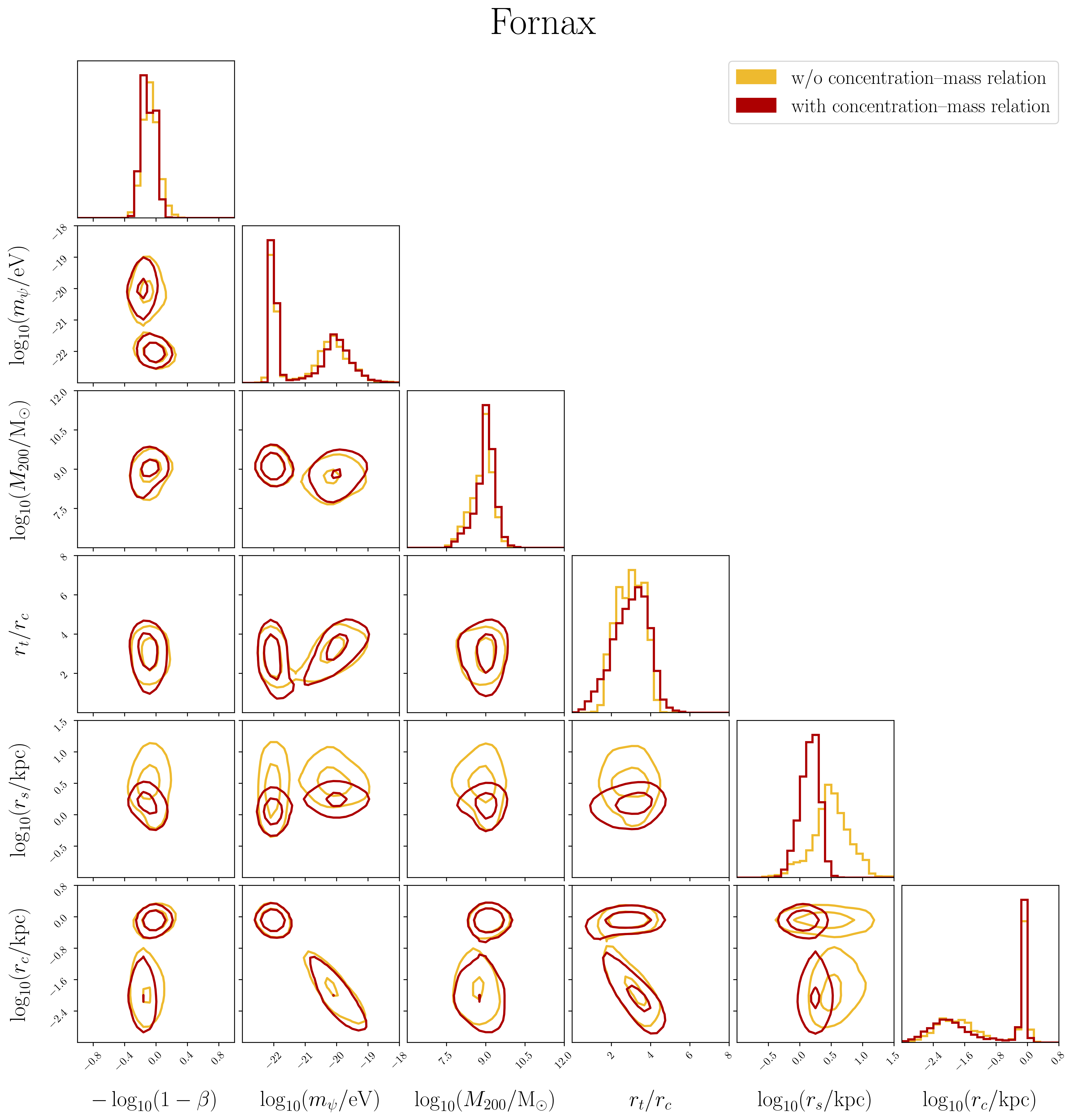}{0.45\textwidth}{}
\fig{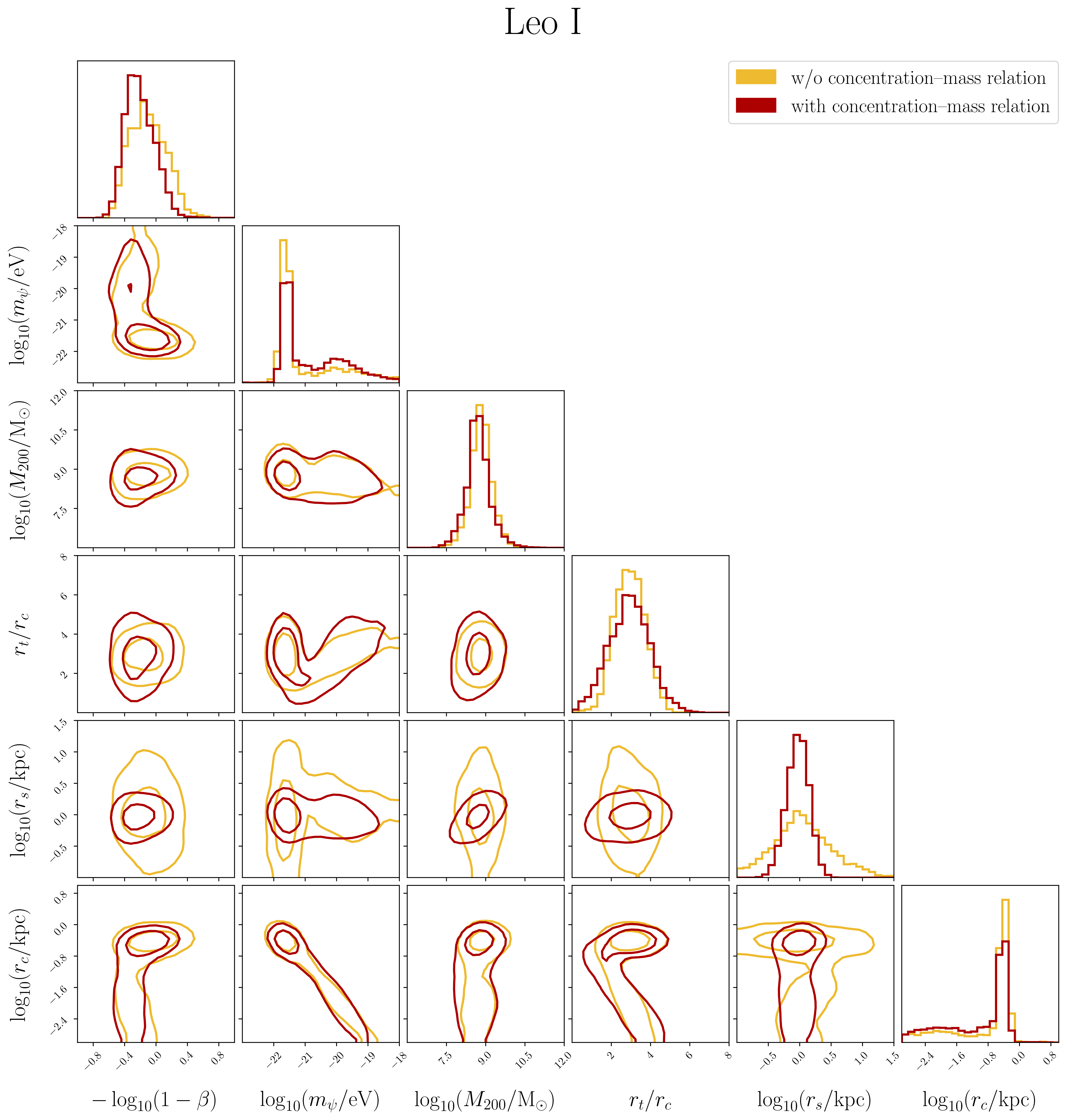}{0.45\textwidth}{}
}
\caption{Posterior distributions of the model parameters for Carina, Draco, Fornax, and Leo~I obtained from models with (red) and without (yellow) the concentration--mass relation.}
\label{fig:posterior_distributions1}
\end{figure*}

\begin{figure*}
\centering
\gridline{
\fig{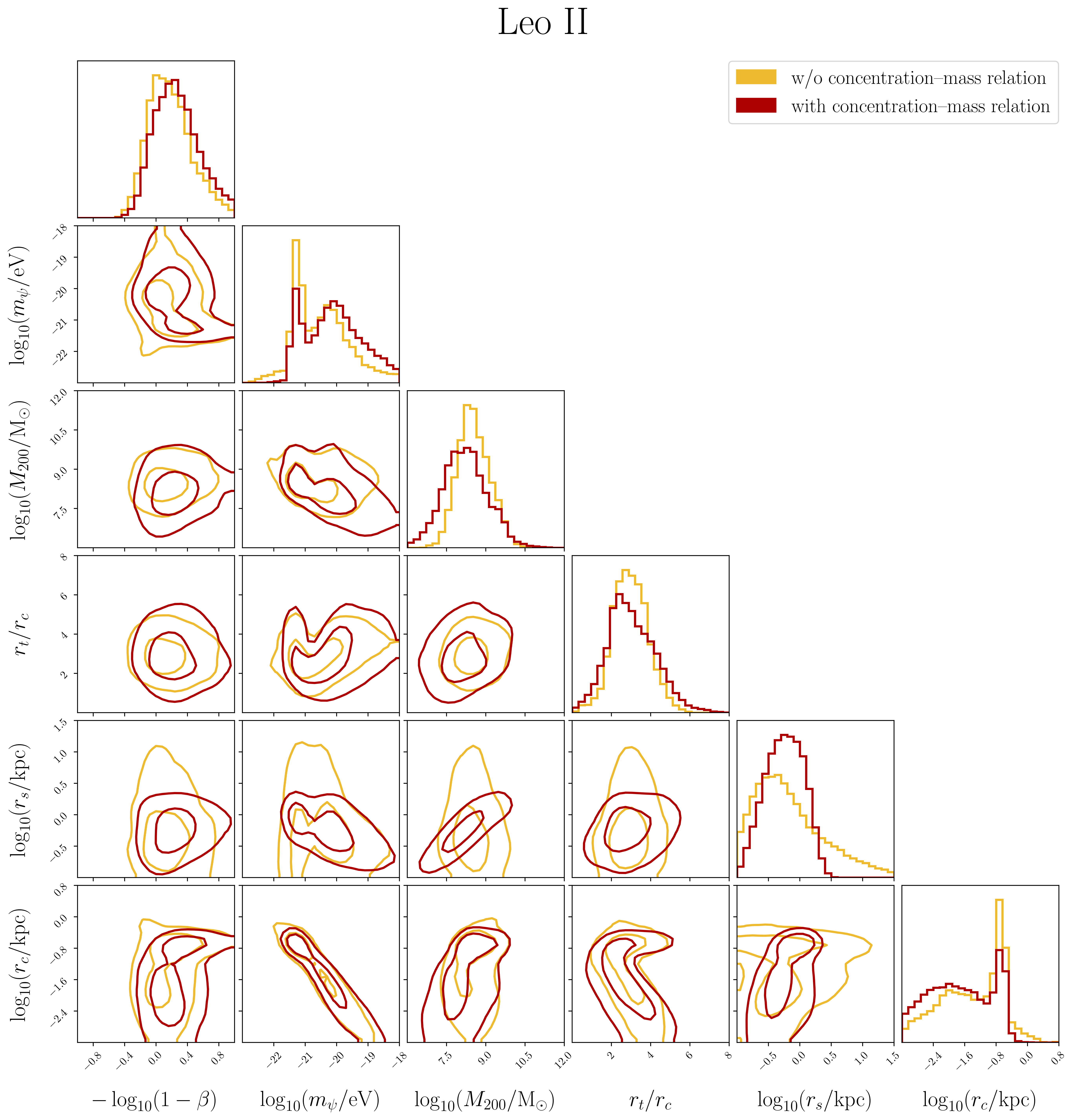}{0.45\textwidth}{}
\fig{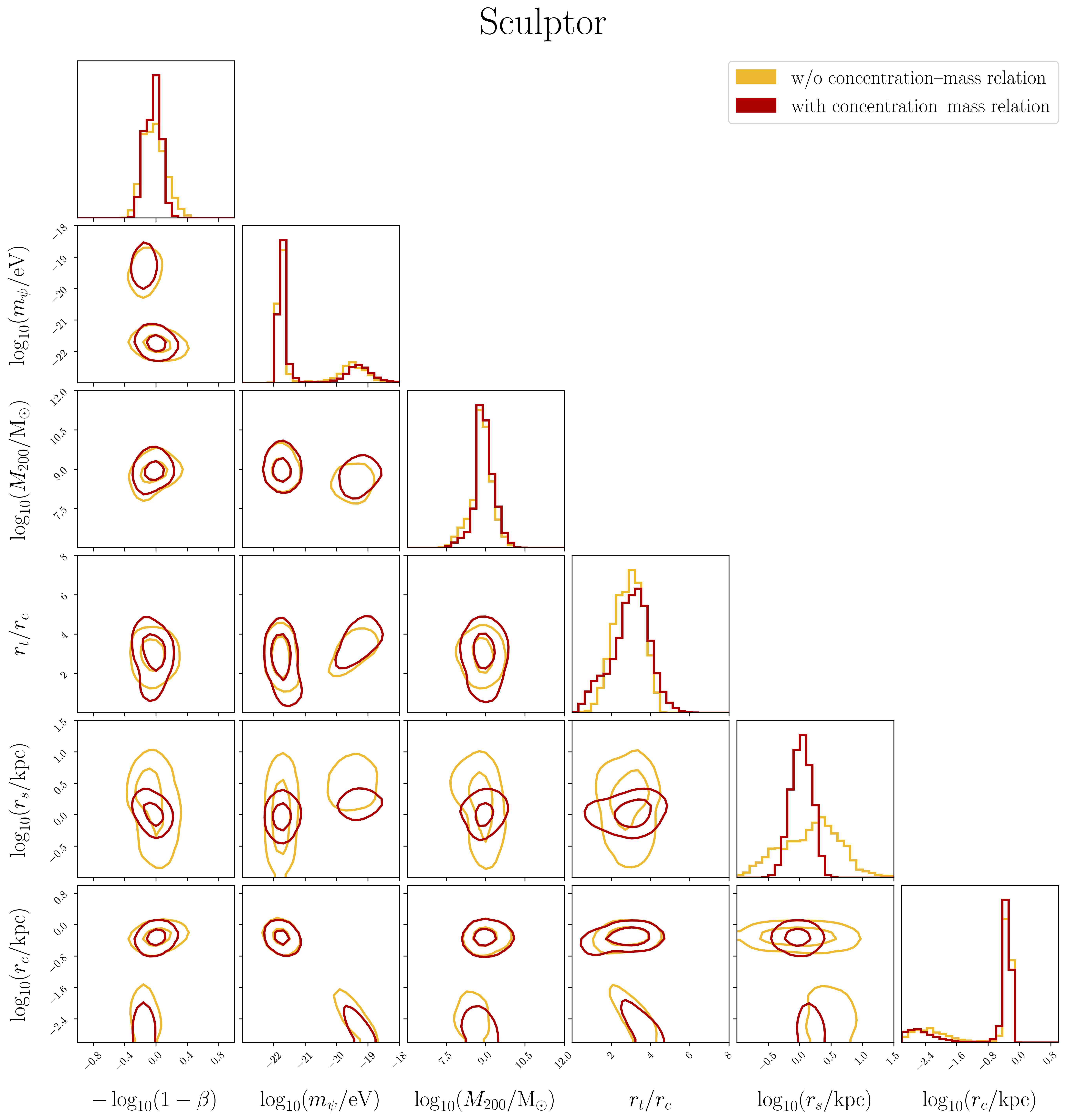}{0.45\textwidth}{}
}
\gridline{
\fig{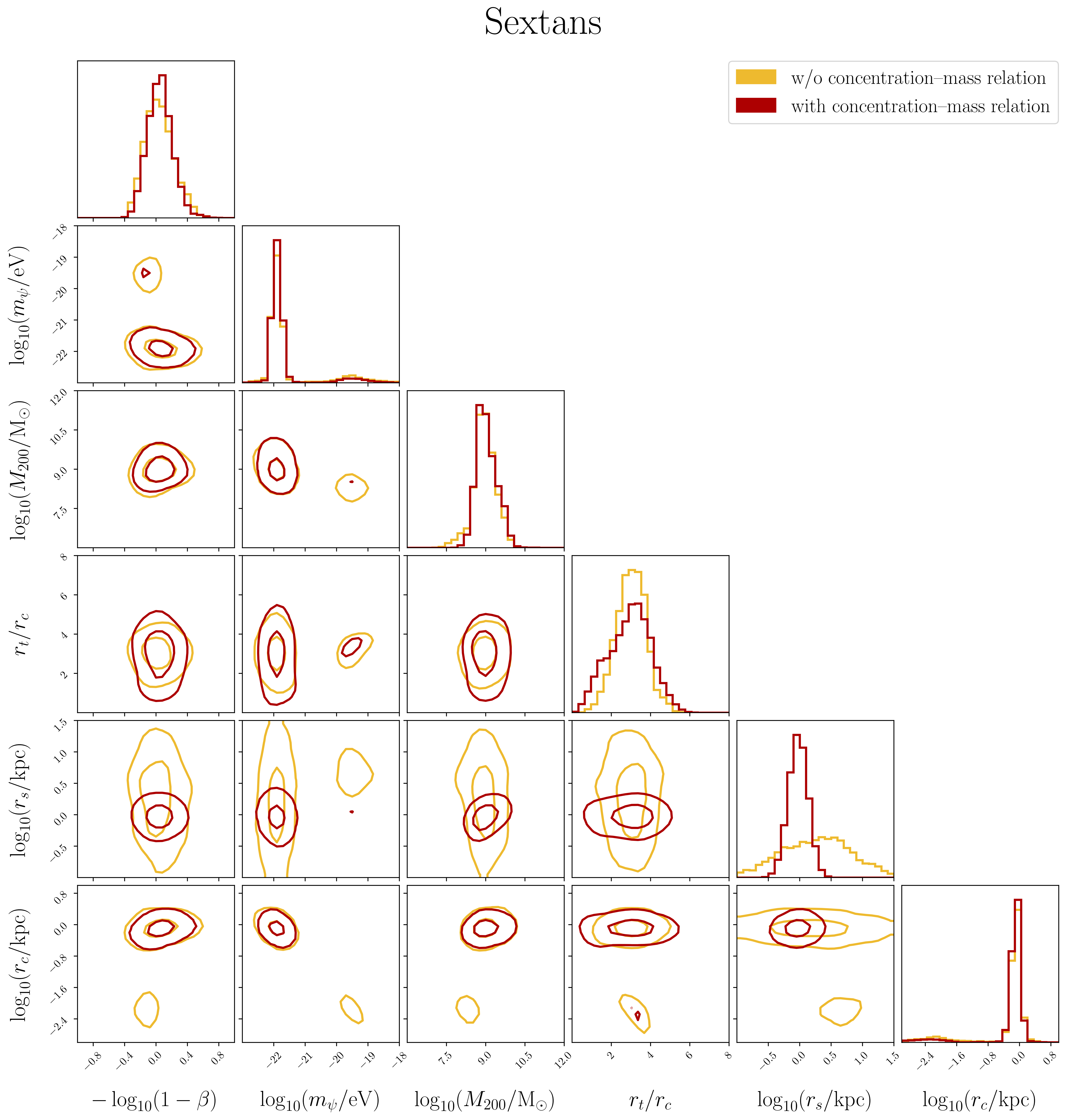}{0.45\textwidth}{}
\fig{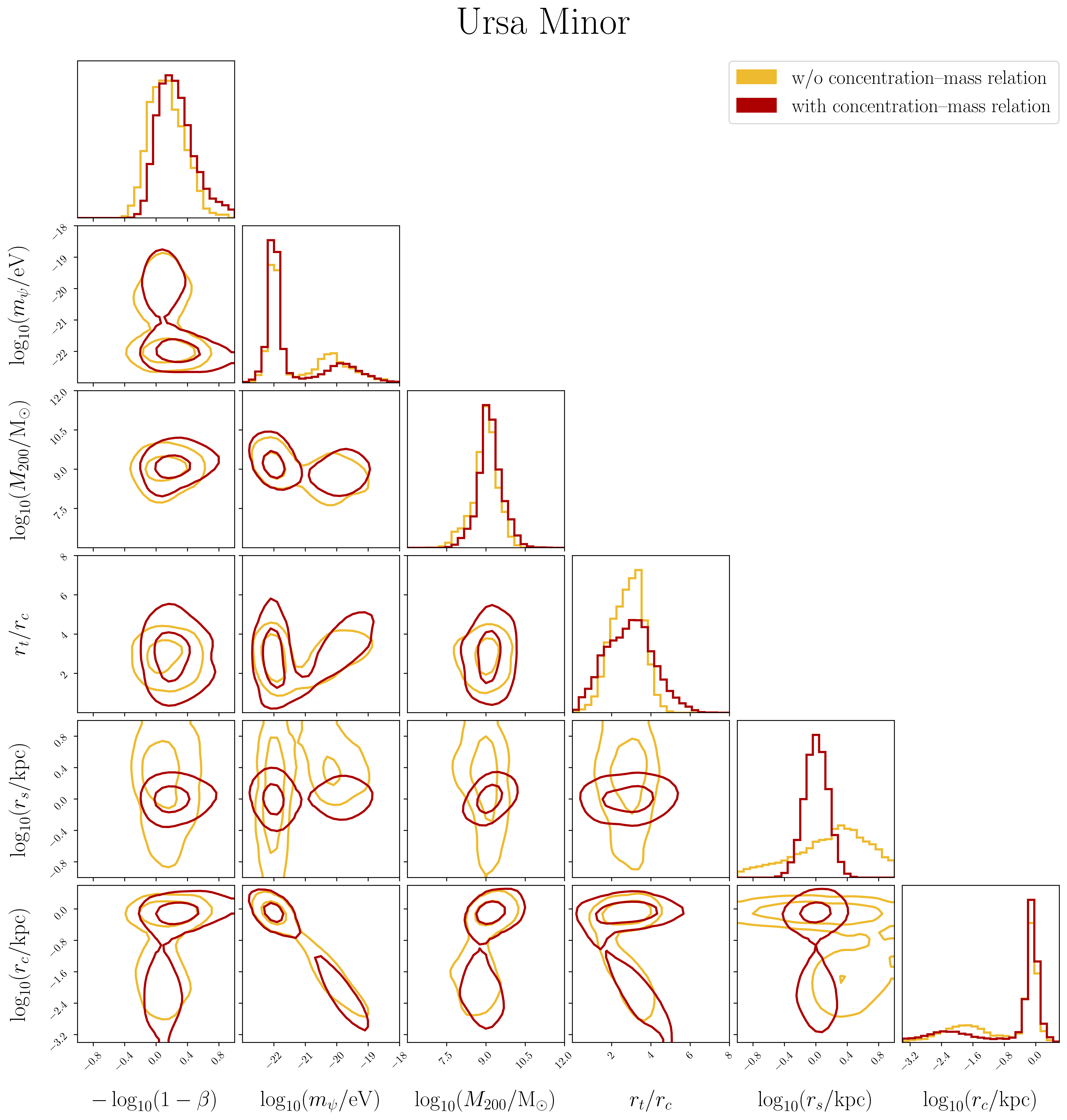}{0.45\textwidth}{}
}
\caption{The same as Figure \ref{fig:posterior_distributions1}, but for Leo II, Sculptor, Sextans, and Ursa Minor.}
\label{fig:posterior_distributions2}
\end{figure*}

We present the full posterior distributions of all model parameters for each analyzed galaxy in Figures~\ref{fig:posterior_distributions1} and \ref{fig:posterior_distributions2}. For reference, we also include the inferred core radius $r_c$, although it is not treated as a free parameter in the model.

\begin{figure*}
  \centering
  \includegraphics[width=0.8\textwidth]{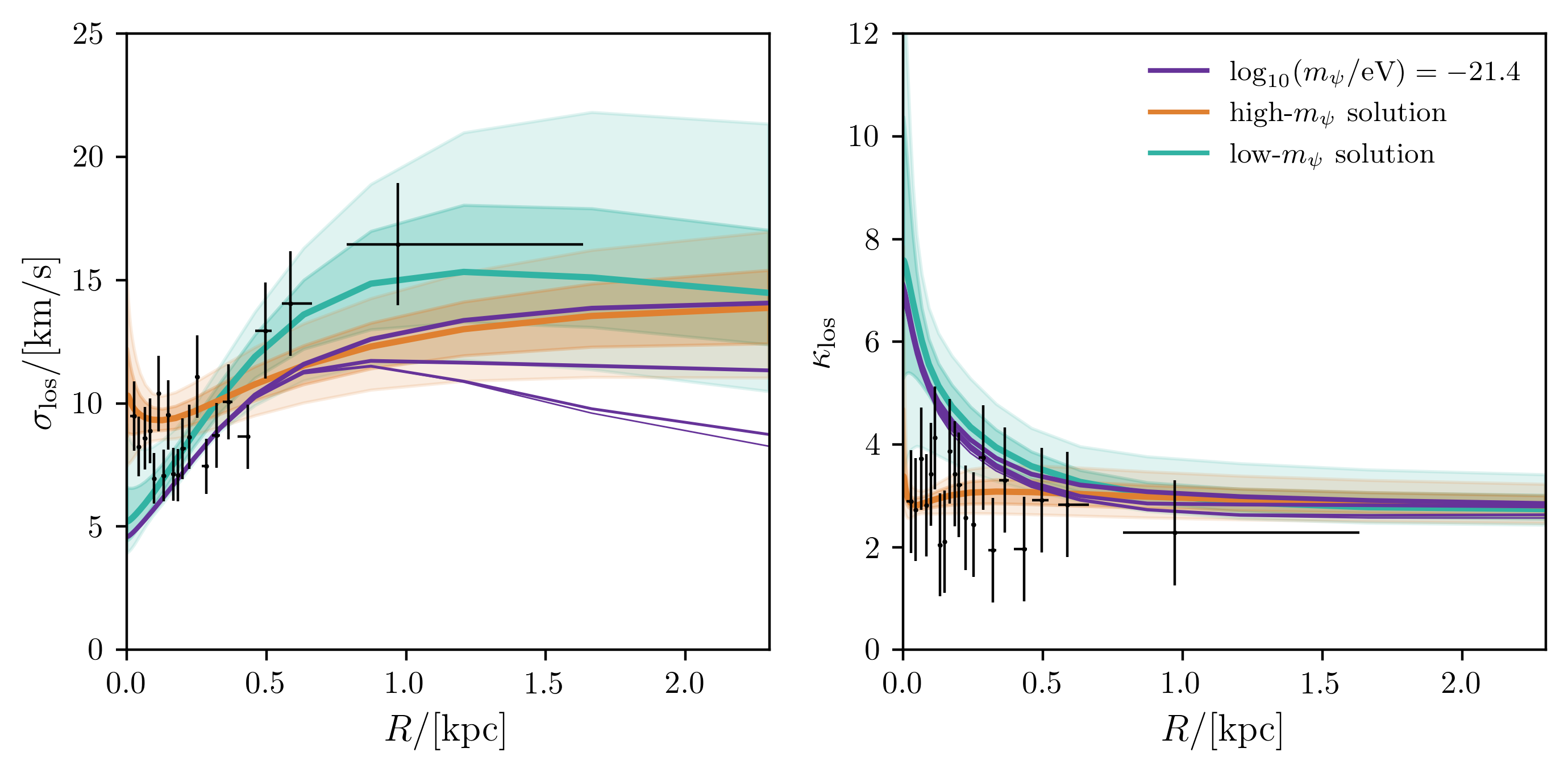}
  \caption{The left and right panels show the recovered LOS velocity dispersion and LOS kurtosis profiles of the Draco dSph, respectively. The darker and lighter shaded areas indicate the 68\% and 95\% credible intervals, with green and orange corresponding to the low- and high-$m_\psi$ solutions. The purple curves show the predicted $\sigma_{\textrm{los}}$ and $\kappa_{\textrm{los}}$ profiles for $\log_{10}(m_\psi/\textrm{eV}) = -21.4$. Different line thicknesses correspond to $r_t/r_c = [1, 2, 3, 4]$, ordered from the thickest to the thinnest. All other model parameters are fixed to the median values of their posterior distributions. Black points with error bars denote the observed data, with uncertainties corresponding to the 68\% credible intervals.}
  \label{fig:kinematicsrecovery}
\end{figure*}

A comparison between the model predictions and the observed kinematic data is shown in Figure~\ref{fig:kinematicsrecovery}. The left and right panels display the recovered LOS velocity dispersion and LOS kurtosis profiles for the Draco dSph, respectively. The darker and lighter shaded bands represent the 68\% and 95\% credible intervals, with green and orange corresponding to the low- and high-$m_\psi$ solutions. The black data points with error bars denote the observed measurements. Although the likelihood analysis is performed using individual stellar measurements without binning, the data are binned in this figure for visualization purposes. The purple curves illustrate model kinematic profiles computed for a fixed $\log_{10}(m_\psi/\textrm{eV}) = -21.4$, while varying the transition radius as $r_t/r_c = [1,2,3,4]$, ordered from the thickest to the thinnest line. These curves illustrate why this range of $m_\psi$ is disfavored by the model.

\section{Gelman--Rubin statistic} \label{sec:GelmanRubin}
The Gelman--Rubin statistic, $\hat{R}$, is a convergence diagnostic in MCMC analyses. It assesses whether multiple chains have converged to the same target distribution by comparing the variance within individual chains to the variance between chains. The statistic is defined as
\begin{equation}
    \hat{R} = \sqrt{\frac{\frac{N_{\rm s}-1}{N_{\rm s}}W + \frac{1}{N_{\rm s}}B}{W}}.
\end{equation}
The average within-chain variance, $W$, is given by
\begin{equation}
    W = \frac{1}{N_{\rm chain}} \sum_{j=1}^{N_{\rm chain}} s_j^2,
\end{equation}
where $s_j^2$ represents the sample variance of the $j$-th chain,
\begin{equation}
    s_j^2 = \frac{1}{N_{\rm s}-1} \sum_{i=1}^{N_{\rm s}} \left( \theta_{ij} - \bar{\theta}_j \right)^2.
\end{equation}
The between-chain variance, $B$, denotes the dispersion among the chain means and is defined as
\begin{equation}
    B = \frac{N_{\rm s}}{N_{\rm chain}-1} \sum_{j=1}^{N_{\rm chain}} \left( \bar{\theta}_j - \bar{\theta} \right)^2,
\end{equation}
where
\begin{equation}
\bar{\theta}j = \frac{1}{N{\rm s}} \sum_{i=1}^{N_{\rm s}} \theta_{ij}
\end{equation}
is the mean of the $j$-th chain, and
\begin{equation}
\bar{\theta} = \frac{1}{N_{\rm chain}} \sum_{j=1}^{N_{\rm chain}} \bar{\theta}_j
\end{equation}
is the mean across all chains. Here, $N_{\rm chain}$ denotes the number of Markov chains, $N_{\rm s}$ is the number of samples in each chain, and $\theta_{ij}$ represents the $i$-th sample of the $j$-th chain.

If Markov chains have converged, they are likely sampling from the same posterior distribution and are statistically indistinguishable from one another. Thus, the between-chain variance should be about the same as the within-chain variance, and $\hat{R}$ should be close to unity.

\bibliography{sample7}{}

@ARTICLE{Banares2023,
       author = {{Ba{\~n}ares-Hern{\'a}ndez}, Andr{\'e}s and {Castillo}, Andr{\'e}s and {Martin Camalich}, Jorge and {Iorio}, Giuliano},
        title = "{Confronting fuzzy dark matter with the rotation curves of nearby dwarf irregular galaxies}",
      journal = {\aap},
         year = 2023,
        month = aug,
       volume = {676},
          eid = {A63},
        pages = {A63},
          doi = {10.1051/0004-6361/202346686},
archivePrefix = {arXiv},
       eprint = {2304.05793},
 primaryClass = {astro-ph.GA},
       adsurl = {https://ui.adsabs.harvard.edu/abs/2023A&A...676A..63B}
}

@ARTICLE{Banares2026,
       author = {{Ba{\~n}ares-Hern{\'a}ndez}, Andr{\'e}s and {Read}, Justin I. and {J{\'u}lio}, Mariana P.},
        title = "{GravSphere2: A higher order Jeans method for mass modeling spherical stellar systems}",
      journal = {\aap},
         year = 2026,
        month = jan,
       volume = {705},
          eid = {A212},
        pages = {A212},
          doi = {10.1051/0004-6361/202557195},
archivePrefix = {arXiv},
       eprint = {2509.24103},
 primaryClass = {astro-ph.GA},
       adsurl = {https://ui.adsabs.harvard.edu/abs/2026A&A...705A.212B}
}

@ARTICLE{Banik2021,
       author = {{Banik}, Nilanjan and {Bovy}, Jo and {Bertone}, Gianfranco and {Erkal}, Denis and {de Boer}, T.~J.~L.},
        title = "{Novel constraints on the particle nature of dark matter from stellar streams}",
      journal = {\jcap},
         year = 2021,
        month = oct,
       volume = {2021},
       number = {10},
          eid = {043},
        pages = {043},
          doi = {10.1088/1475-7516/2021/10/043},
archivePrefix = {arXiv},
       eprint = {1911.02663},
 primaryClass = {astro-ph.GA},
       adsurl = {https://ui.adsabs.harvard.edu/abs/2021JCAP...10..043B}
}

@ARTICLE{Bar2018,
       author = {{Bar}, Nitsan and {Blas}, Diego and {Blum}, Kfir and {Sibiryakov}, Sergey},
        title = "{Galactic rotation curves versus ultralight dark matter: Implications of the soliton-host halo relation}",
      journal = {\prd},
         year = 2018,
        month = oct,
       volume = {98},
       number = {8},
          eid = {083027},
        pages = {083027},
          doi = {10.1103/PhysRevD.98.083027},
archivePrefix = {arXiv},
       eprint = {1805.00122},
 primaryClass = {astro-ph.CO},
       adsurl = {https://ui.adsabs.harvard.edu/abs/2018PhRvD..98h3027B}
}

@ARTICLE{Bar2022,
       author = {{Bar}, Nitsan and {Blum}, Kfir and {Sun}, Chen},
        title = "{Galactic rotation curves versus ultralight dark matter: A systematic comparison with SPARC data}",
      journal = {\prd},
         year = 2022,
        month = apr,
       volume = {105},
       number = {8},
          eid = {083015},
        pages = {083015},
          doi = {10.1103/PhysRevD.105.083015},
archivePrefix = {arXiv},
       eprint = {2111.03070},
 primaryClass = {hep-ph},
       adsurl = {https://ui.adsabs.harvard.edu/abs/2022PhRvD.105h3015B}
}

@ARTICLE{Battaglia2013,
       author = {{Battaglia}, Giuseppina and {Helmi}, Amina and {Breddels}, Maarten},
        title = "{Internal kinematics and dynamical models of dwarf spheroidal galaxies around the Milky Way}",
      journal = {\nar},
         year = 2013,
        month = sep,
       volume = {57},
       number = {3-4},
        pages = {52-79},
          doi = {10.1016/j.newar.2013.05.003},
archivePrefix = {arXiv},
       eprint = {1305.5965},
 primaryClass = {astro-ph.CO},
       adsurl = {https://ui.adsabs.harvard.edu/abs/2013NewAR..57...52B}
}

@ARTICLE{BattagliaNipoti2022,
       author = {{Battaglia}, Giuseppina and {Nipoti}, Carlo},
        title = "{Stellar dynamics and dark matter in Local Group dwarf galaxies}",
      journal = {Nature Astronomy},
         year = 2022,
        month = may,
       volume = {6},
        pages = {659-672},
          doi = {10.1038/s41550-022-01638-7},
archivePrefix = {arXiv},
       eprint = {2205.07821},
 primaryClass = {astro-ph.GA},
       adsurl = {https://ui.adsabs.harvard.edu/abs/2022NatAs...6..659B}
}

@ARTICLE{Bellazzini2004,
       author = {{Bellazzini}, M. and {Gennari}, N. and {Ferraro}, F.~R. and {Sollima}, A.},
        title = "{The distance to the Leo I dwarf spheroidal galaxy from the red giant branch tip}",
      journal = {Monthly Notices of the Royal Astronomical Society},
         year = 2004,
        month = nov,
       volume = {354},
       number = {3},
        pages = {708-712},
          doi = {10.1111/j.1365-2966.2004.08226.x},
archivePrefix = {arXiv},
       eprint = {astro-ph/0407444},
 primaryClass = {astro-ph},
       adsurl = {https://ui.adsabs.harvard.edu/abs/2004MNRAS.354..708B}
}

@ARTICLE{Bellazzini2005,
       author = {{Bellazzini}, M. and {Gennari}, N. and {Ferraro}, F.~R.},
        title = "{The red giant branch tip and bump of the Leo II dwarf spheroidal galaxy}",
      journal = {\mnras},
         year = 2005,
        month = jun,
       volume = {360},
       number = {1},
        pages = {185-193},
          doi = {10.1111/j.1365-2966.2005.09027.x},
archivePrefix = {arXiv},
       eprint = {astro-ph/0503418},
 primaryClass = {astro-ph},
       adsurl = {https://ui.adsabs.harvard.edu/abs/2005MNRAS.360..185B}
}

@ARTICLE{Benito2020,
       author = {{Benito}, Mar{\'\i}a and {Criado}, Juan Carlos and {H{\"u}tsi}, Gert and {Raidal}, Martti and {Veerm{\"a}e}, Hardi},
        title = "{Implications of Milky Way substructures for the nature of dark matter}",
      journal = {\prd},
         year = 2020,
        month = may,
       volume = {101},
       number = {10},
          eid = {103023},
        pages = {103023},
          doi = {10.1103/PhysRevD.101.103023},
archivePrefix = {arXiv},
       eprint = {2001.11013},
 primaryClass = {astro-ph.CO},
       adsurl = {https://ui.adsabs.harvard.edu/abs/2020PhRvD.101j3023B}
}

@ARTICLE{Bernal2018,
       author = {{Bernal}, T. and {Fern{\'a}ndez-Hern{\'a}ndez}, L.~M. and {Matos}, T. and {Rodr{\'\i}guez-Meza}, M.~A.},
        title = "{Rotation curves of high-resolution LSB and SPARC galaxies with fuzzy and multistate (ultralight boson) scalar field dark matter}",
      journal = {\mnras},
         year = 2018,
        month = apr,
       volume = {475},
       number = {2},
        pages = {1447-1468},
          doi = {10.1093/mnras/stx3208},
archivePrefix = {arXiv},
       eprint = {1701.00912},
 primaryClass = {astro-ph.GA},
       adsurl = {https://ui.adsabs.harvard.edu/abs/2018MNRAS.475.1447B}
}

@BOOK{BinneyMerrifield1998,
       author = {{Binney}, James and {Merrifield}, Michael},
        title = "{Galactic Astronomy}",
         year = 1998,
       adsurl = {https://ui.adsabs.harvard.edu/abs/1998gaas.book.....B},
    publisher = {Princeton University Press}
}

@BOOK{BinneyTremaine2008,
       author = {{Binney}, James and {Tremaine}, Scott},
        title = "{Galactic Dynamics: Second Edition}",
         year = 2008,
       adsurl = {https://ui.adsabs.harvard.edu/abs/2008gady.book.....B},
    publisher = {Princeton University Press}
}

@ARTICLE{Bonanos2004,
       author = {{Bonanos}, A.~Z. and {Stanek}, K.~Z. and {Szentgyorgyi}, A.~H. and {Sasselov}, D.~D. and {Bakos}, G. {\'A}.},
        title = "{The RR Lyrae Distance to the Draco Dwarf Spheroidal Galaxy}",
      journal = {The Astronomical Journal},
         year = 2004,
        month = feb,
       volume = {127},
       number = {2},
        pages = {861-867},
          doi = {10.1086/381073},
archivePrefix = {arXiv},
       eprint = {astro-ph/0310477},
 primaryClass = {astro-ph},
       adsurl = {https://ui.adsabs.harvard.edu/abs/2004AJ....127..861B}
}

@ARTICLE{BullockBoylan-Kolchin2017,
       author = {{Bullock}, James S. and {Boylan-Kolchin}, Michael},
        title = "{Small-Scale Challenges to the {\ensuremath{\Lambda}}CDM Paradigm}",
      journal = {Annual Review of Astronomy and Astrophysics},
         year = 2017,
        month = aug,
       volume = {55},
       number = {1},
        pages = {343-387},
          doi = {10.1146/annurev-astro-091916-055313},
archivePrefix = {arXiv},
       eprint = {1707.04256},
 primaryClass = {astro-ph.CO},
       adsurl = {https://ui.adsabs.harvard.edu/abs/2017ARA&A..55..343B}
}

@ARTICLE{Carrera2002,
       author = {{Carrera}, Ricardo and {Aparicio}, Antonio and {Mart{\'\i}nez-Delgado}, David and {Alonso-Garc{\'\i}a}, Javier},
        title = "{The Star Formation History and Spatial Distribution of Stellar Populations in the Ursa Minor Dwarf Spheroidal Galaxy}",
      journal = {The Astronomical Journal},
         year = 2002,
        month = jun,
       volume = {123},
       number = {6},
        pages = {3199-3209},
          doi = {10.1086/340702},
archivePrefix = {arXiv},
       eprint = {astro-ph/0203300},
 primaryClass = {astro-ph},
       adsurl = {https://ui.adsabs.harvard.edu/abs/2002AJ....123.3199C}
}

@ARTICLE{Chan2021,
       author = {{Chan}, Man Ho and {Fai Yeung}, Chu},
        title = "{Model-independent Constraints on Ultralight Dark Matter from the SPARC Data}",
      journal = {\apj},
         year = 2021,
        month = may,
       volume = {913},
       number = {1},
          eid = {25},
        pages = {25},
          doi = {10.3847/1538-4357/abf42f},
archivePrefix = {arXiv},
       eprint = {2104.05159},
 primaryClass = {astro-ph.GA},
       adsurl = {https://ui.adsabs.harvard.edu/abs/2021ApJ...913...25C}
}

@ARTICLE{Chema2023,
       author = {{Arroyo-Polonio}, Jos{\'e} Mar{\'\i}a and {Battaglia}, Giuseppina and {Thomas}, Guillaume F. and {Irwin}, Michael J. and {McConnachie}, Alan W. and {Tolstoy}, Eline},
        title = "{Binary star population of the Sculptor dwarf galaxy}",
      journal = {\aap},
         year = 2023,
        month = sep,
       volume = {677},
          eid = {A95},
        pages = {A95},
          doi = {10.1051/0004-6361/202346843},
archivePrefix = {arXiv},
       eprint = {2307.10375},
 primaryClass = {astro-ph.GA},
       adsurl = {https://ui.adsabs.harvard.edu/abs/2023A&A...677A..95A}
}

@ARTICLE{Chema2026,
       author = {{Mar{\'\i}a Arroyo-Polonio}, Jos{\'e} and {Battaglia}, Giuseppina and {Thomas}, Guillaume F.},
        title = "{Estimating the dynamical masses of dwarf galaxies in the presence of binary-star contamination}",
      journal = {arXiv e-prints},
         year = 2026,
        month = mar,
          eid = {arXiv:2603.03129},
        pages = {arXiv:2603.03129},
          doi = {10.48550/arXiv.2603.03129},
archivePrefix = {arXiv},
       eprint = {2603.03129},
 primaryClass = {astro-ph.GA},
       adsurl = {https://ui.adsabs.harvard.edu/abs/2026arXiv260303129M}
}

@ARTICLE{Chen2017,
       author = {{Chen}, Shu-Rong and {Schive}, Hsi-Yu and {Chiueh}, Tzihong},
        title = "{Jeans analysis for dwarf spheroidal galaxies in wave dark matter}",
      journal = {\mnras},
         year = 2017,
        month = jun,
       volume = {468},
       number = {2},
        pages = {1338-1348},
          doi = {10.1093/mnras/stx449},
archivePrefix = {arXiv},
       eprint = {1606.09030},
 primaryClass = {astro-ph.GA},
       adsurl = {https://ui.adsabs.harvard.edu/abs/2017MNRAS.468.1338C}
}

@ARTICLE{Chiba2026,
       author = {{Chiba}, Masashi and {Wyse}, Rosemary F.~G. and {Kirby}, Evan N. and {Cohen}, Judith G. and {Dobos}, L{\'a}szl{\'o} and {Gerasimov}, Roman and {Ishigaki}, Miho N. and {Hayashi}, Kohei and {Filion}, Carrie and {Arnaboldi}, Magda and {Bhattacharya}, Souradeep and {Hirai}, Yutaka and {Kobayashi}, Chiaki and {Komiyama}, Yutaka and {Kuzma}, Pete B. and {Ogami}, Itsuki and {Chies-Santos}, Ana L. and {Klock-Miranda}, Nicole L. and {Sestito}, Federico and {Budav{\'a}ri}, Tam{\'a}s and {Cooper}, Andrew P. and {Ding}, Keyi and {Escala}, Ivanna and {Ferreira}, Elisa G.~M. and {Gerhard}, Ortwin and {Henderson}, Lauren and {Hong}, Jihye and {Horigome}, Shunichi and {Ikeda}, Ryota and {Ishikawa}, Ryo and {Kirihara}, Takanobu and {Li}, Zhuohan and {Martin}, Nicolas and {Miyazaki Sakurako Okamoto}, Rin and {Pattnaik}, Rohan and {Sato}, Kyosuke and {Suzuki}, Yoshihisa and {Szalay}, Alexander S. and {Wardana}, Dafa and {Wei}, Viska and {Wu}, Wenbo and {Wu}, Zhenyu and {Xu}, Xinfeng and {Ye}, Xianhao and {Miki}, Yohei and {Zhang}, Xiangwei and {Zhao}, Gang and {Zhao}, Jingkun and {Zhao}, Xiaosheng},
        title = "{Galactic Archaeology with the Subaru `{\={O}}nohi`ula Prime Focus Spectrograph Strategic Program}",
      journal = {arXiv e-prints},
         year = 2026,
        month = apr,
          eid = {arXiv:2604.09875},
        pages = {arXiv:2604.09875},
          doi = {10.48550/arXiv.2604.09875},
archivePrefix = {arXiv},
       eprint = {2604.09875},
 primaryClass = {astro-ph.GA},
       adsurl = {https://ui.adsabs.harvard.edu/abs/2026arXiv260409875C}
}

@ARTICLE{Chowdhury2021,
       author = {{Dutta Chowdhury}, Dhruba and {van den Bosch}, Frank C. and {Robles}, Victor H. and {van Dokkum}, Pieter and {Schive}, Hsi-Yu and {Chiueh}, Tzihong and {Broadhurst}, Tom},
        title = "{On the Random Motion of Nuclear Objects in a Fuzzy Dark Matter Halo}",
      journal = {\apj},
         year = 2021,
        month = jul,
       volume = {916},
       number = {1},
          eid = {27},
        pages = {27},
          doi = {10.3847/1538-4357/ac043f},
archivePrefix = {arXiv},
       eprint = {2105.05268},
 primaryClass = {astro-ph.GA},
       adsurl = {https://ui.adsabs.harvard.edu/abs/2021ApJ...916...27D}
}

@ARTICLE{Cooper2023,
       author = {{Cooper}, Andrew P. and {Koposov}, Sergey E. and {Allende Prieto}, Carlos and {Manser}, Christopher J. and {Kizhuprakkat}, Namitha and {Myers}, Adam D. and {Dey}, Arjun and {G{\"a}nsicke}, Boris T. and {Li}, Ting S. and {Rockosi}, Constance and {Valluri}, Monica and {Najita}, Joan and {Deason}, Alis and {Raichoor}, Anand and {Wang}, M.-Y. and {Ting}, Y.-S. and {Kim}, Bokyoung and {Carrillo}, Andreia and {Wang}, Wenting and {Beraldo e Silva}, Leandro and {Han}, Jiwon Jesse and {Ding}, Jiani and {S{\'a}nchez-Conde}, Miguel and {Aguilar}, Jessica N. and {Ahlen}, Steven and {Bailey}, Stephen and {Belokurov}, Vasily and {Brooks}, David and {Cunha}, Katia and {Dawson}, Kyle and {de la Macorra}, Axel and {Doel}, Peter and {Eisenstein}, Daniel J. and {Fagrelius}, Parker and {Fanning}, Kevin and {Font-Ribera}, Andreu and {Forero-Romero}, Jaime E. and {Gazta{\~n}aga}, Enrique and {Gontcho a Gontcho}, Satya and {Guy}, Julien and {Honscheid}, Klaus and {Kehoe}, Robert and {Kisner}, Theodore and {Kremin}, Anthony and {Landriau}, Martin and {Levi}, Michael E. and {Martini}, Paul and {Meisner}, Aaron M. and {Miquel}, Ramon and {Moustakas}, John and {Nie}, Jundan J.~D. and {Palanque-Delabrouille}, Nathalie and {Percival}, Will J. and {Poppett}, Claire and {Prada}, Francisco and {Rehemtulla}, Nabeel and {Schlafly}, Edward and {Schlegel}, David and {Schubnell}, Michael and {Sharples}, Ray M. and {Tarl{\'e}}, Gregory and {Wechsler}, Risa H. and {Weinberg}, David H. and {Zhou}, Zhimin and {Zou}, Hu},
        title = "{Overview of the DESI Milky Way Survey}",
      journal = {\apj},
         year = 2023,
        month = apr,
       volume = {947},
       number = {1},
          eid = {37},
        pages = {37},
          doi = {10.3847/1538-4357/acb3c0},
archivePrefix = {arXiv},
       eprint = {2208.08514},
 primaryClass = {astro-ph.GA},
       adsurl = {https://ui.adsabs.harvard.edu/abs/2023ApJ...947...37C}
}

@ARTICLE{Dalal2022,
       author = {{Dalal}, Neal and {Kravtsov}, Andrey},
        title = "{Excluding fuzzy dark matter with sizes and stellar kinematics of ultrafaint dwarf galaxies}",
      journal = {Physical Review D},
         year = 2022,
        month = sep,
       volume = {106},
       number = {6},
          eid = {063517},
        pages = {063517},
          doi = {10.1103/PhysRevD.106.063517},
archivePrefix = {arXiv},
       eprint = {2203.05750},
 primaryClass = {astro-ph.CO},
       adsurl = {https://ui.adsabs.harvard.edu/abs/2022PhRvD.106f3517D}
}

@INPROCEEDINGS{Dalton2012,
       author = {{Dalton}, Gavin and {Trager}, Scott C. and {Abrams}, Don Carlos and {Carter}, David and {Bonifacio}, Piercarlo and {Aguerri}, J. Alfonso L. and {MacIntosh}, Mike and {Evans}, Chris and {Lewis}, Ian and {Navarro}, Ramon and {Agocs}, Tibor and {Dee}, Kevin and {Rousset}, Sophie and {Tosh}, Ian and {Middleton}, Kevin and {Pragt}, Johannes and {Terrett}, David and {Brock}, Matthew and {Benn}, Chris and {Verheijen}, Marc and {Cano Infantes}, Diego and {Bevil}, Craige and {Steele}, Iain and {Mottram}, Chris and {Bates}, Stuart and {Gribbin}, Francis J. and {Rey}, J{\"u}rg and {Rodriguez}, Luis Fernando and {Delgado}, Jose Miguel and {Guinouard}, Isabelle and {Walton}, Nic and {Irwin}, Michael J. and {Jagourel}, Pascal and {Stuik}, Remko and {Gerlofsma}, Gerrit and {Roelfsma}, Ronald and {Skillen}, Ian and {Ridings}, Andy and {Balcells}, Marc and {Daban}, Jean-Baptiste and {Gouvret}, Carole and {Venema}, Lars and {Girard}, Paul},
        title = "{WEAVE: the next generation wide-field spectroscopy facility for the William Herschel Telescope}",
    booktitle = {Ground-based and Airborne Instrumentation for Astronomy IV},
         year = 2012,
       editor = {{McLean}, Ian S. and {Ramsay}, Suzanne K. and {Takami}, Hideki},
       series = {Society of Photo-Optical Instrumentation Engineers (SPIE) Conference Series},
       volume = {8446},
        month = sep,
          eid = {84460P},
        pages = {84460P},
          doi = {10.1117/12.925950},
       adsurl = {https://ui.adsabs.harvard.edu/abs/2012SPIE.8446E..0PD}
}

@ARTICLE{Davoudiasl2019,
       author = {{Davoudiasl}, Hooman and {Denton}, Peter B.},
        title = "{Ultralight Boson Dark Matter and Event Horizon Telescope Observations of M 87$^{*}$}",
      journal = {\prl},
         year = 2019,
        month = jul,
       volume = {123},
       number = {2},
          eid = {021102},
        pages = {021102},
          doi = {10.1103/PhysRevLett.123.021102},
archivePrefix = {arXiv},
       eprint = {1904.09242},
 primaryClass = {astro-ph.CO},
       adsurl = {https://ui.adsabs.harvard.edu/abs/2019PhRvL.123b1102D}
}

@ARTICLE{Eberhardt2025b,
       author = {{Eberhardt}, Andrew and {Ferreira}, Elisa G.~M.},
        title = "{Ultralight fuzzy dark matter review}",
      journal = {arXiv e-prints},
         year = 2025,
        month = jul,
          eid = {arXiv:2507.00705},
        pages = {arXiv:2507.00705},
          doi = {10.48550/arXiv.2507.00705},
archivePrefix = {arXiv},
       eprint = {2507.00705},
 primaryClass = {astro-ph.CO},
       adsurl = {https://ui.adsabs.harvard.edu/abs/2025arXiv250700705E}
}

@ARTICLE{Eberhardt2025c,
       author = {{Eberhardt}, Andrew and {Gosenca}, Mateja and {Hui}, Lam},
        title = "{Heating and scattering of stellar distributions by ultralight dark matter}",
      journal = {arXiv e-prints},
         year = 2025,
        month = oct,
          eid = {arXiv:2510.17079},
        pages = {arXiv:2510.17079},
          doi = {10.48550/arXiv.2510.17079},
archivePrefix = {arXiv},
       eprint = {2510.17079},
 primaryClass = {astro-ph.CO},
       adsurl = {https://ui.adsabs.harvard.edu/abs/2025arXiv251017079E}
}

@ARTICLE{Farisy2025,
       author = {{Al Farisy}, Fahmi M. and {Wulandari}, Hesti R.~T. and {Dante}, Azriel J.},
        title = "{Properties of Standard, Fuzzy, and Self-interacting Dark Matter Haloes in Dwarf Galaxies}",
      journal = {Research in Astronomy and Astrophysics},
         year = 2025,
        month = apr,
       volume = {25},
       number = {4},
          eid = {045006},
        pages = {045006},
          doi = {10.1088/1674-4527/adc29c},
       adsurl = {https://ui.adsabs.harvard.edu/abs/2025RAA....25d5006A}
}

@ARTICLE{Fabrizio2016,
       author = {{Fabrizio}, M. and {Bono}, G. and {Nonino}, M. and {{\L}okas}, E.~L. and {Ferraro}, I. and {Iannicola}, G. and {Buonanno}, R. and {Cassisi}, S. and {Coppola}, G. and {Dall'Ora}, M. and {Gilmozzi}, R. and {Marconi}, M. and {Monelli}, M. and {Romaniello}, M. and {Stetson}, P.~B. and {Th{\'e}venin}, F. and {Walker}, A.~R.},
        title = "{The Carina Project. X. On the Kinematics of Old and Intermediate-age Stellar Populations1,2}",
      journal = {The Astrophysical Journal},
         year = 2016,
        month = oct,
       volume = {830},
       number = {2},
          eid = {126},
        pages = {126},
          doi = {10.3847/0004-637X/830/2/126},
archivePrefix = {arXiv},
       eprint = {1607.03181},
 primaryClass = {astro-ph.GA},
       adsurl = {https://ui.adsabs.harvard.edu/abs/2016ApJ...830..126F}
}

@ARTICLE{Ferreira2021,
       author = {{Ferreira}, Elisa G.~M.},
        title = "{Ultra-light dark matter}",
      journal = {Astronomy \& Astrophysics Review},
         year = 2021,
        month = dec,
       volume = {29},
       number = {1},
          eid = {7},
        pages = {7},
          doi = {10.1007/s00159-021-00135-6},
archivePrefix = {arXiv},
       eprint = {2005.03254},
 primaryClass = {astro-ph.CO},
       adsurl = {https://ui.adsabs.harvard.edu/abs/2021A&ARv..29....7F}
}

@ARTICLE{Gelman1992,
       author = {{Gelman}, Andrew and {Rubin}, Donald B.},
        title = "{Inference from Iterative Simulation Using Multiple Sequences}",
      journal = {Statistical Science},
         year = 1992,
        month = jan,
       volume = {7},
        pages = {457-472},
          doi = {10.1214/ss/1177011136},
       adsurl = {https://ui.adsabs.harvard.edu/abs/1992StaSc...7..457G}
}

@ARTICLE{Genina2020,
       author = {{Genina}, A. and {Read}, J.~I. and {Frenk}, C.~S. and {Cole}, S. and {Ben{\'\i}tez-Llambay}, A. and {Ludlow}, A.~D. and {Navarro}, J.~F. and {Oman}, K.~A. and {Robertson}, A.},
        title = "{To {\ensuremath{\beta}} or not to {\ensuremath{\beta}}: can higher order Jeans analysis break the mass-anisotropy degeneracy in simulated dwarfs?}",
      journal = {\mnras},
         year = 2020,
        month = oct,
       volume = {498},
       number = {1},
        pages = {144-163},
          doi = {10.1093/mnras/staa2352},
archivePrefix = {arXiv},
       eprint = {1911.09124},
 primaryClass = {astro-ph.GA},
       adsurl = {https://ui.adsabs.harvard.edu/abs/2020MNRAS.498..144G}
}

@ARTICLE{Goldstein2022,
       author = {{Goldstein}, Isabelle S. and {Koushiappas}, Savvas M. and {Walker}, Matthew G.},
        title = "{Viability of ultralight bosonic dark matter in dwarf galaxies}",
      journal = {\prd},
         year = 2022,
        month = sep,
       volume = {106},
       number = {6},
          eid = {063010},
        pages = {063010},
          doi = {10.1103/PhysRevD.106.063010},
archivePrefix = {arXiv},
       eprint = {2206.05244},
 primaryClass = {astro-ph.GA},
       adsurl = {https://ui.adsabs.harvard.edu/abs/2022PhRvD.106f3010G}
}

@ARTICLE{GonzalezMorales2017,
       author = {{Gonz{\'a}lez-Morales}, Alma X. and {Marsh}, David J.~E. and {Pe{\~n}arrubia}, Jorge and {Ure{\~n}a-L{\'o}pez}, Luis A.},
        title = "{Unbiased constraints on ultralight axion mass from dwarf spheroidal galaxies}",
      journal = {\mnras},
         year = 2017,
        month = dec,
       volume = {472},
       number = {2},
        pages = {1346-1360},
          doi = {10.1093/mnras/stx1941},
archivePrefix = {arXiv},
       eprint = {1609.05856},
 primaryClass = {astro-ph.CO},
       adsurl = {https://ui.adsabs.harvard.edu/abs/2017MNRAS.472.1346G}
}

@ARTICLE{Hastings1970,
       author = {{Hastings}, W.~K.},
        title = "{Monte Carlo Sampling Methods using Markov Chains and their Applications}",
      journal = {Biometrika},
         year = 1970,
        month = apr,
       volume = {57},
       number = {1},
        pages = {97-109},
          doi = {10.1093/biomet/57.1.97},
       adsurl = {https://ui.adsabs.harvard.edu/abs/1970Bimka..57...97H}
}

@ARTICLE{HayashiObata2020,
       author = {{Hayashi}, Kohei and {Obata}, Ippei},
        title = "{Non-sphericity of ultralight-axion dark matter haloes in the Galactic dwarf spheroidal galaxies}",
      journal = {\mnras},
         year = 2020,
        month = jan,
       volume = {491},
       number = {1},
        pages = {615-633},
          doi = {10.1093/mnras/stz2950},
archivePrefix = {arXiv},
       eprint = {1902.03054},
 primaryClass = {astro-ph.CO},
       adsurl = {https://ui.adsabs.harvard.edu/abs/2020MNRAS.491..615H}
}

@ARTICLE{Hayashi2021,
       author = {{Hayashi}, Kohei and {Ferreira}, Elisa G.~M. and {Chan}, Hei Yin Jowett},
        title = "{Narrowing the Mass Range of Fuzzy Dark Matter with Ultrafaint Dwarfs}",
      journal = {\apjl},
         year = 2021,
        month = may,
       volume = {912},
       number = {1},
          eid = {L3},
        pages = {L3},
          doi = {10.3847/2041-8213/abf501},
archivePrefix = {arXiv},
       eprint = {2102.05300},
 primaryClass = {astro-ph.CO},
       adsurl = {https://ui.adsabs.harvard.edu/abs/2021ApJ...912L...3H}
}

@ARTICLE{Hlozek2015,
       author = {{Hlozek}, Ren{\'e}e and {Grin}, Daniel and {Marsh}, David J.~E. and {Ferreira}, Pedro G.},
        title = "{A search for ultralight axions using precision cosmological data}",
      journal = {Physical Review D},
         year = 2015,
        month = may,
       volume = {91},
       number = {10},
          eid = {103512},
        pages = {103512},
          doi = {10.1103/PhysRevD.91.103512},
archivePrefix = {arXiv},
       eprint = {1410.2896},
 primaryClass = {astro-ph.CO},
       adsurl = {https://ui.adsabs.harvard.edu/abs/2015PhRvD..91j3512H}
}

@ARTICLE{Hlozek2018,
       author = {{Hlo{\v{z}}ek}, Ren{\'e}e and {Marsh}, David J.~E. and {Grin}, Daniel},
        title = "{Using the full power of the cosmic microwave background to probe axion dark matter}",
      journal = {Monthly Notices of the Royal Astronomical Society},
         year = 2018,
        month = may,
       volume = {476},
       number = {3},
        pages = {3063-3085},
          doi = {10.1093/mnras/sty271},
archivePrefix = {arXiv},
       eprint = {1708.05681},
 primaryClass = {astro-ph.CO},
       adsurl = {https://ui.adsabs.harvard.edu/abs/2018MNRAS.476.3063H}
}

@article{Horigome2026,
  author = {{Ando}, Shin'ichiro and {Ferreira}, Elisa G.~M. and {Horigome}, Shunichi},
  title  = {Cosmology-informed constraints on the fuzzy dark matter mass from dwarf-spheroidal stellar kinematics},
  note   = {to appear},
  year   = {2026}
}

@ARTICLE{Hu2000,
       author = {{Hu}, Wayne and {Barkana}, Rennan and {Gruzinov}, Andrei},
        title = "{Fuzzy Cold Dark Matter: The Wave Properties of Ultralight Particles}",
      journal = {Physical Review Letters},
         year = 2000,
        month = aug,
       volume = {85},
       number = {6},
        pages = {1158-1161},
          doi = {10.1103/PhysRevLett.85.1158},
archivePrefix = {arXiv},
       eprint = {astro-ph/0003365},
 primaryClass = {astro-ph},
       adsurl = {https://ui.adsabs.harvard.edu/abs/2000PhRvL..85.1158H}
}

@ARTICLE{Hui2017,
       author = {{Hui}, Lam and {Ostriker}, Jeremiah P. and {Tremaine}, Scott and {Witten}, Edward},
        title = "{Ultralight scalars as cosmological dark matter}",
      journal = {\prd},
         year = 2017,
        month = feb,
       volume = {95},
       number = {4},
          eid = {043541},
        pages = {043541},
          doi = {10.1103/PhysRevD.95.043541},
archivePrefix = {arXiv},
       eprint = {1610.08297},
 primaryClass = {astro-ph.CO},
       adsurl = {https://ui.adsabs.harvard.edu/abs/2017PhRvD..95d3541H}
}

@ARTICLE{Hui2021,
       author = {{Hui}, Lam},
        title = "{Wave Dark Matter}",
      journal = {\araa},
         year = 2021,
        month = sep,
       volume = {59},
        pages = {247-289},
          doi = {10.1146/annurev-astro-120920-010024},
archivePrefix = {arXiv},
       eprint = {2101.11735},
 primaryClass = {astro-ph.CO},
       adsurl = {https://ui.adsabs.harvard.edu/abs/2021ARA&A..59..247H}
}

@ARTICLE{Hunter2007,
       author = {{Hunter}, John D.},
        title = "{Matplotlib: A 2D Graphics Environment}",
      journal = {Computing in Science and Engineering},
         year = 2007,
        month = jan,
       volume = {9},
       number = {3},
        pages = {90-95},
          doi = {10.1109/MCSE.2007.55},
       adsurl = {https://ui.adsabs.harvard.edu/abs/2007CSE.....9...90H}
}

@ARTICLE{Irsic2017,
       author = {{Ir{\v{s}}i{\v{c}}}, Vid and {Viel}, Matteo and {Haehnelt}, Martin G. and {Bolton}, James S. and {Becker}, George D.},
        title = "{First Constraints on Fuzzy Dark Matter from Lyman-{\ensuremath{\alpha}} Forest Data and Hydrodynamical Simulations}",
      journal = {\prl},
         year = 2017,
        month = jul,
       volume = {119},
       number = {3},
          eid = {031302},
        pages = {031302},
          doi = {10.1103/PhysRevLett.119.031302},
archivePrefix = {arXiv},
       eprint = {1703.04683},
 primaryClass = {astro-ph.CO},
       adsurl = {https://ui.adsabs.harvard.edu/abs/2017PhRvL.119c1302I}
}

@ARTICLE{Ishiyama2021,
       author = {{Ishiyama}, Tomoaki and {Prada}, Francisco and {Klypin}, Anatoly A. and {Sinha}, Manodeep and {Metcalf}, R. Benton and {Jullo}, Eric and {Altieri}, Bruno and {Cora}, Sof{\'\i}a A. and {Croton}, Darren and {de la Torre}, Sylvain and {Mill{\'a}n-Calero}, David E. and {Oogi}, Taira and {Ruedas}, Jos{\'e} and {Vega-Mart{\'\i}nez}, Cristian A.},
        title = "{The Uchuu simulations: Data Release 1 and dark matter halo concentrations}",
      journal = {\mnras},
         year = 2021,
        month = sep,
       volume = {506},
       number = {3},
        pages = {4210-4231},
          doi = {10.1093/mnras/stab1755},
archivePrefix = {arXiv},
       eprint = {2007.14720},
 primaryClass = {astro-ph.CO},
       adsurl = {https://ui.adsabs.harvard.edu/abs/2021MNRAS.506.4210I}
}

@ARTICLE{Jin2024,
       author = {{Jin}, Shoko and {Trager}, Scott C. and {Dalton}, Gavin B. and {Aguerri}, J. Alfonso L. and {Drew}, J.~E. and {Falc{\'o}n-Barroso}, Jes{\'u}s and {G{\"a}nsicke}, Boris T. and {Hill}, Vanessa and {Iovino}, Angela and {Pieri}, Matthew M. and {Poggianti}, Bianca M. and {Smith}, D.~J.~B. and {Vallenari}, Antonella and {Abrams}, Don Carlos and {Aguado}, David S. and {Antoja}, Teresa and {Arag{\'o}n-Salamanca}, Alfonso and {Ascasibar}, Yago and {Babusiaux}, Carine and {Balcells}, Marc and {Barrena}, R. and {Battaglia}, Giuseppina and {Belokurov}, Vasily and {Bensby}, Thomas and {Bonifacio}, Piercarlo and {Bragaglia}, Angela and {Carrasco}, Esperanza and {Carrera}, Ricardo and {Cornwell}, Daniel J. and {Dom{\'\i}nguez-Palmero}, Lilian and {Duncan}, Kenneth J. and {Famaey}, Benoit and {Fari{\~n}a}, Cecilia and {Gonzalez}, Oscar A. and {Guest}, Steve and {Hatch}, Nina A. and {Hess}, Kelley M. and {Hoskin}, Matthew J. and {Irwin}, Mike and {Knapen}, Johan H. and {Koposov}, Sergey E. and {Kuchner}, Ulrike and {Laigle}, Clotilde and {Lewis}, Jim and {Longhetti}, Marcella and {Lucatello}, Sara and {M{\'e}ndez-Abreu}, Jairo and {Mercurio}, Amata and {Molaeinezhad}, Alireza and {Mongui{\'o}}, Maria and {Morrison}, Sean and {Murphy}, David N.~A. and {Peralta de Arriba}, Luis and {P{\'e}rez}, Isabel and {P{\'e}rez-R{\`a}fols}, Ignasi and {Pic{\'o}}, Sergio and {Raddi}, Roberto and {Romero-G{\'o}mez}, Merc{\`e} and {Royer}, Fr{\'e}d{\'e}ric and {Siebert}, Arnaud and {Seabroke}, George M. and {Som}, Debopam and {Terrett}, David and {Thomas}, Guillaume and {Wesson}, Roger and {Worley}, C. Clare and {Alfaro}, Emilio J. and {Allende Prieto}, Carlos and {Alonso-Santiago}, Javier and {Amos}, Nicholas J. and {Ashley}, Richard P. and {Balaguer-N{\'u}{\~n}ez}, Lola and {Balbinot}, Eduardo and {Bellazzini}, Michele and {Benn}, Chris R. and {Berlanas}, Sara R. and {Bernard}, Edouard J. and {Best}, Philip and {Bettoni}, Daniela and {Bianco}, Andrea and {Bishop}, Georgia and {Blomqvist}, Michael and {Boeche}, Corrado and {Bolzonella}, Micol and {Bonoli}, Silvia and {Bosma}, Albert and {Britavskiy}, Nikolay and {Busarello}, Gianni and {Caffau}, Elisabetta and {Cantat-Gaudin}, Tristan and {Castro-Ginard}, Alfred and {Couto}, Guilherme and {Carbajo-Hijarrubia}, Juan and {Carter}, David and {Casamiquela}, Laia and {Conrado}, Ana M. and {Corcho-Caballero}, Pablo and {Costantin}, Luca and {Deason}, Alis and {de Burgos}, Abel and {De Grandi}, Sabrina and {Di Matteo}, Paola and {Dom{\'\i}nguez-G{\'o}mez}, Jes{\'u}s and {Dorda}, Ricardo and {Drake}, Alyssa and {Dutta}, Rajeshwari and {Erkal}, Denis and {Feltzing}, Sofia and {Ferr{\'e}-Mateu}, Anna and {Feuillet}, Diane and {Figueras}, Francesca and {Fossati}, Matteo and {Franciosini}, Elena and {Frasca}, Antonio and {Fumagalli}, Michele and {Gallazzi}, Anna and {Garc{\'\i}a-Benito}, Rub{\'e}n and {Gentile Fusillo}, Nicola and {Gebran}, Marwan and {Gilbert}, James and {Gledhill}, T.~M. and {Gonz{\'a}lez Delgado}, Rosa M. and {Greimel}, Robert and {Guarcello}, Mario Giuseppe and {Guerra}, Jose and {Gullieuszik}, Marco and {Haines}, Christopher P. and {Hardcastle}, Martin J. and {Harris}, Amy and {Haywood}, Misha and {Helmi}, Amina and {Hernandez}, Nauzet and {Herrero}, Artemio and {Hughes}, Sarah and {Ir{\v{s}}i{\v{c}}}, Vid and {Jablonka}, Pascale and {Jarvis}, Matt J. and {Jordi}, Carme and {Kondapally}, Rohit and {Kordopatis}, Georges and {Krogager}, Jens-Kristian and {La Barbera}, Francesco and {Lam}, Man I. and {Larsen}, S{\o}ren S. and {Lemasle}, Bertrand and {Lewis}, Ian J. and {Lhom{\'e}}, Emilie and {Lind}, Karin and {Lodi}, Marcello and {Longobardi}, Alessia and {Lonoce}, Ilaria and {Magrini}, Laura and {Ma{\'\i}z Apell{\'a}niz}, Jes{\'u}s and {Marchal}, Olivier and {Marco}, Amparo and {Martin}, Nicolas F. and {Matsuno}, Tadafumi and {Maurogordato}, Sophie and {Merluzzi}, Paola and {Miralda-Escud{\'e}}, Jordi and {Molinari}, Emilio and {Monari}, Giacomo and {Morelli}, Lorenzo and {Mottram}, Christopher J. and {Naylor}, Tim and {Negueruela}, Ignacio and {O{\~n}orbe}, Jose and {Pancino}, Elena and {Peirani}, S{\'e}bastien and {Peletier}, Reynier F. and {Pozzetti}, Lucia and {Rainer}, Monica and {Ramos}, Pau and {Read}, Shaun C. and {Rossi}, Elena Maria and {R{\"o}ttgering}, Huub J.~A. and {Rubi{\~n}o-Mart{\'\i}n}, Jose Alberto and {Sabater}, Jose and {San Juan}, Jos{\'e} and {Sanna}, Nicoletta and {Schallig}, Ellen and {Schiavon}, Ricardo P. and {Schultheis}, Mathias and {Serra}, Paolo and {Shimwell}, Timothy W. and {Sim{\'o}n-D{\'\i}az}, Sergio and {Smith}, Russell J. and {Sordo}, Rosanna and {Sorini}, Daniele and {Soubiran}, Caroline and {Starkenburg}, Else and {Steele}, Iain A. and {Stott}, John and {Stuik}, Remko and {Tolstoy}, Eline and {Tortora}, Crescenzo and {Tsantaki}, Maria and {Van der Swaelmen}, Mathieu and {van Weeren}, Reinout J. and {Vergani}, Daniela},
        title = "{The wide-field, multiplexed, spectroscopic facility WEAVE: Survey design, overview, and simulated implementation}",
      journal = {\mnras},
         year = 2024,
        month = may,
       volume = {530},
       number = {3},
        pages = {2688-2730},
          doi = {10.1093/mnras/stad557},
archivePrefix = {arXiv},
       eprint = {2212.03981},
 primaryClass = {astro-ph.IM},
       adsurl = {https://ui.adsabs.harvard.edu/abs/2024MNRAS.530.2688J}
}

@ARTICLE{Jowett2022,
       author = {{Chan}, Hei Yin Jowett and {Ferreira}, Elisa G.~M. and {May}, Simon and {Hayashi}, Kohei and {Chiba}, Masashi},
        title = "{The diversity of core-halo structure in the fuzzy dark matter model}",
      journal = {\mnras},
         year = 2022,
        month = mar,
       volume = {511},
       number = {1},
        pages = {943-952},
          doi = {10.1093/mnras/stac063},
archivePrefix = {arXiv},
       eprint = {2110.11882},
 primaryClass = {astro-ph.CO},
       adsurl = {https://ui.adsabs.harvard.edu/abs/2022MNRAS.511..943C}
}

@ARTICLE{Kawai2024,
       author = {{Kawai}, Hiroki and {Kamada}, Ayuki and {Kamada}, Kohei and {Yoshida}, Naoki},
        title = "{Modeling the core-halo mass relation in fuzzy dark matter halos}",
      journal = {\prd},
         year = 2024,
        month = jul,
       volume = {110},
       number = {2},
          eid = {023519},
        pages = {023519},
          doi = {10.1103/PhysRevD.110.023519},
archivePrefix = {arXiv},
       eprint = {2312.10744},
 primaryClass = {astro-ph.CO},
       adsurl = {https://ui.adsabs.harvard.edu/abs/2024PhRvD.110b3519K}
}

@ARTICLE{Khelashvili2023,
       author = {{Khelashvili}, M. and {Rudakovskyi}, A. and {Hossenfelder}, S.},
        title = "{Dark matter profiles of SPARC galaxies: a challenge to fuzzy dark matter}",
      journal = {\mnras},
         year = 2023,
        month = aug,
       volume = {523},
       number = {3},
        pages = {3393-3405},
          doi = {10.1093/mnras/stad1595},
archivePrefix = {arXiv},
       eprint = {2207.14165},
 primaryClass = {astro-ph.CO},
       adsurl = {https://ui.adsabs.harvard.edu/abs/2023MNRAS.523.3393K}
}

@ARTICLE{Kirby2013,
       author = {{Kirby}, Evan N. and {Boylan-Kolchin}, Michael and {Cohen}, Judith G. and {Geha}, Marla and {Bullock}, James S. and {Kaplinghat}, Manoj},
        title = "{Segue 2: The Least Massive Galaxy}",
      journal = {\apj},
         year = 2013,
        month = jun,
       volume = {770},
       number = {1},
          eid = {16},
        pages = {16},
          doi = {10.1088/0004-637X/770/1/16},
archivePrefix = {arXiv},
       eprint = {1304.6080},
 primaryClass = {astro-ph.CO},
       adsurl = {https://ui.adsabs.harvard.edu/abs/2013ApJ...770...16K}
}

@ARTICLE{Kirby2017,
       author = {{Kirby}, Evan N. and {Cohen}, Judith G. and {Simon}, Joshua D. and {Guhathakurta}, Puragra and {Thygesen}, Anders O. and {Duggan}, Gina E.},
        title = "{Triangulum II. Not Especially Dense After All}",
      journal = {\apj},
         year = 2017,
        month = apr,
       volume = {838},
       number = {2},
          eid = {83},
        pages = {83},
          doi = {10.3847/1538-4357/aa6570},
archivePrefix = {arXiv},
       eprint = {1703.02978},
 primaryClass = {astro-ph.GA},
       adsurl = {https://ui.adsabs.harvard.edu/abs/2017ApJ...838...83K}
}

@ARTICLE{Koch2007,
       author = {{Koch}, Andreas and {Kleyna}, Jan T. and {Wilkinson}, Mark I. and {Grebel}, Eva K. and {Gilmore}, Gerard F. and {Evans}, N. Wyn and {Wyse}, Rosemary F.~G. and {Harbeck}, Daniel R.},
        title = "{Stellar Kinematics in the Remote Leo II Dwarf Spheroidal Galaxy-Another Brick in the Wall}",
      journal = {The Astronomical Journal},
         year = 2007,
        month = aug,
       volume = {134},
       number = {2},
        pages = {566-578},
          doi = {10.1086/519380},
archivePrefix = {arXiv},
       eprint = {0704.3437},
 primaryClass = {astro-ph},
       adsurl = {https://ui.adsabs.harvard.edu/abs/2007AJ....134..566K}
}

@ARTICLE{Laroche2022,
       author = {{Laroche}, Alexander and {Gilman}, Daniel and {Li}, Xinyu and {Bovy}, Jo and {Du}, Xiaolong},
        title = "{Quantum fluctuations masquerade as haloes: bounds on ultra-light dark matter from quadruply imaged quasars}",
      journal = {\mnras},
         year = 2022,
        month = dec,
       volume = {517},
       number = {2},
        pages = {1867-1883},
          doi = {10.1093/mnras/stac2677},
archivePrefix = {arXiv},
       eprint = {2206.11269},
 primaryClass = {astro-ph.CO},
       adsurl = {https://ui.adsabs.harvard.edu/abs/2022MNRAS.517.1867L}
}

@ARTICLE{Lee2009,
       author = {{Lee}, Myung Gyoon and {Yuk}, In-Soo and {Park}, Hong Soo and {Harris}, Jason and {Zaritsky}, Dennis},
        title = "{Star Formation History and Chemical Evolution of the Sextans Dwarf Spheroidal Galaxy}",
      journal = {The Astrophysical Journal},
         year = 2009,
        month = sep,
       volume = {703},
       number = {1},
        pages = {692-701},
          doi = {10.1088/0004-637X/703/1/692},
archivePrefix = {arXiv},
       eprint = {0907.5102},
 primaryClass = {astro-ph.CO},
       adsurl = {https://ui.adsabs.harvard.edu/abs/2009ApJ...703..692L}
}

@ARTICLE{Liao2025,
       author = {{Liao}, Pin-Yu and {Su}, Guan-Ming and {Schive}, Hsi-Yu and {Kunkel}, Alexander and {Huang}, Hsinhao and {Chiueh}, Tzihong},
        title = "{Deciphering the Soliton-Halo Relation in Fuzzy Dark Matter}",
      journal = {\prl},
         year = 2025,
        month = aug,
       volume = {135},
       number = {6},
          eid = {061002},
        pages = {061002},
          doi = {10.1103/9dqj-q6mt},
archivePrefix = {arXiv},
       eprint = {2412.09908},
 primaryClass = {astro-ph.CO},
       adsurl = {https://ui.adsabs.harvard.edu/abs/2025PhRvL.135f1002L}
}

@ARTICLE{Lokas2002,
       author = {{{\L}okas}, Ewa L.},
        title = "{Dark matter distribution in dwarf spheroidal galaxies}",
      journal = {\mnras},
         year = 2002,
        month = jul,
       volume = {333},
       number = {3},
        pages = {697-708},
          doi = {10.1046/j.1365-8711.2002.05457.x},
archivePrefix = {arXiv},
       eprint = {astro-ph/0112023},
 primaryClass = {astro-ph},
       adsurl = {https://ui.adsabs.harvard.edu/abs/2002MNRAS.333..697L}
}

@ARTICLE{Maccio2012,
       author = {{Macci{\`o}}, Andrea V. and {Paduroiu}, Sinziana and {Anderhalden}, Donnino and {Schneider}, Aurel and {Moore}, Ben},
        title = "{Cores in warm dark matter haloes: a Catch 22 problem}",
      journal = {\mnras},
         year = 2012,
        month = aug,
       volume = {424},
       number = {2},
        pages = {1105-1112},
          doi = {10.1111/j.1365-2966.2012.21284.x},
archivePrefix = {arXiv},
       eprint = {1202.1282},
 primaryClass = {astro-ph.CO},
       adsurl = {https://ui.adsabs.harvard.edu/abs/2012MNRAS.424.1105M}
}

@ARTICLE{Marsh2019,
       author = {{Marsh}, David J.~E. and {Niemeyer}, Jens C.},
        title = "{Strong Constraints on Fuzzy Dark Matter from Ultrafaint Dwarf Galaxy Eridanus II}",
      journal = {\prl},
         year = 2019,
        month = aug,
       volume = {123},
       number = {5},
          eid = {051103},
        pages = {051103},
          doi = {10.1103/PhysRevLett.123.051103},
archivePrefix = {arXiv},
       eprint = {1810.08543},
 primaryClass = {astro-ph.CO},
       adsurl = {https://ui.adsabs.harvard.edu/abs/2019PhRvL.123e1103M}
}

@ARTICLE{Martino2023,
       author = {{de Martino}, Ivan},
        title = "{Constraining ultralight bosons in dwarf spheroidal galaxies with a radially varying anisotropy}",
      journal = {\prd},
         year = 2023,
        month = dec,
       volume = {108},
       number = {12},
          eid = {123044},
        pages = {123044},
          doi = {10.1103/PhysRevD.108.123044},
archivePrefix = {arXiv},
       eprint = {2312.07217},
 primaryClass = {astro-ph.GA},
       adsurl = {https://ui.adsabs.harvard.edu/abs/2023PhRvD.108l3044D}
}

@ARTICLE{Mateo2008,
       author = {{Mateo}, Mario and {Olszewski}, Edward W. and {Walker}, Matthew G.},
        title = "{The Velocity Dispersion Profile of the Remote Dwarf Spheroidal Galaxy Leo I: A Tidal Hit and Run?}",
      journal = {The Astrophysical Journal},
         year = 2008,
        month = mar,
       volume = {675},
       number = {1},
        pages = {201-233},
          doi = {10.1086/522326},
archivePrefix = {arXiv},
       eprint = {0708.1327},
 primaryClass = {astro-ph},
       adsurl = {https://ui.adsabs.harvard.edu/abs/2008ApJ...675..201M}
}

@ARTICLE{May2021,
       author = {{May}, Simon and {Springel}, Volker},
        title = "{Structure formation in large-volume cosmological simulations of fuzzy dark matter: impact of the non-linear dynamics}",
      journal = {\mnras},
         year = 2021,
        month = sep,
       volume = {506},
       number = {2},
        pages = {2603-2618},
          doi = {10.1093/mnras/stab1764},
archivePrefix = {arXiv},
       eprint = {2101.01828},
 primaryClass = {astro-ph.CO},
       adsurl = {https://ui.adsabs.harvard.edu/abs/2021MNRAS.506.2603M}
}

@ARTICLE{May2025,
       author = {{May}, Simon and {Dalal}, Neal and {Kravtsov}, Andrey},
        title = "{Updated bounds on ultra-light dark matter from the tiniest galaxies}",
      journal = {arXiv e-prints},
         year = 2025,
        month = sep,
          eid = {arXiv:2509.02781},
        pages = {arXiv:2509.02781},
          doi = {10.48550/arXiv.2509.02781},
archivePrefix = {arXiv},
       eprint = {2509.02781},
 primaryClass = {astro-ph.CO},
       adsurl = {https://ui.adsabs.harvard.edu/abs/2025arXiv250902781M}
}

@ARTICLE{McConnachie2010,
       author = {{McConnachie}, Alan W. and {C{\^o}t{\'e}}, Patrick},
        title = "{Revisiting the Influence of Unidentified Binaries on Velocity Dispersion Measurements in Ultra-faint Stellar Systems}",
      journal = {\apjl},
         year = 2010,
        month = oct,
       volume = {722},
       number = {2},
        pages = {L209-L214},
          doi = {10.1088/2041-8205/722/2/L209},
archivePrefix = {arXiv},
       eprint = {1009.4205},
 primaryClass = {astro-ph.CO},
       adsurl = {https://ui.adsabs.harvard.edu/abs/2010ApJ...722L.209M}
}

@ARTICLE{McConnachie2012,
       author = {{McConnachie}, Alan W.},
        title = "{The Observed Properties of Dwarf Galaxies in and around the Local Group}",
      journal = {The Astronomical Journal},
         year = 2012,
        month = jul,
       volume = {144},
       number = {1},
          eid = {4},
        pages = {4},
          doi = {10.1088/0004-6256/144/1/4},
archivePrefix = {arXiv},
       eprint = {1204.1562},
 primaryClass = {astro-ph.CO},
       adsurl = {https://ui.adsabs.harvard.edu/abs/2012AJ....144....4M}
}

@INPROCEEDINGS{Merrifield1990,
       author = {{Merrifield}, M.~R. and {Kent}, S.~M.},
        title = "{Extracting the Dynamics of Spherical Systems from Higher Order Velocity Moments}",
    booktitle = {Bulletin of the American Astronomical Society},
         year = 1990,
       volume = {22},
        month = jan,
        pages = {744},
       adsurl = {https://ui.adsabs.harvard.edu/abs/1990BAAS...22..744M}
}

@ARTICLE{Metropolis1953,
       author = {{Metropolis}, Nicholas and {Rosenbluth}, Arianna W. and {Rosenbluth}, Marshall N. and {Teller}, Augusta H. and {Teller}, Edward},
        title = "{Equation of State Calculations by Fast Computing Machines}",
      journal = {\jcp},
         year = 1953,
        month = jun,
       volume = {21},
       number = {6},
        pages = {1087-1092},
          doi = {10.1063/1.1699114},
       adsurl = {https://ui.adsabs.harvard.edu/abs/1953JChPh..21.1087M}
}

@ARTICLE{Mina2022,
       author = {{Mina}, Mattia and {Mota}, David F. and {Winther}, Hans A.},
        title = "{Solitons in the dark: First approach to non-linear structure formation with fuzzy dark matter}",
      journal = {\aap},
         year = 2022,
        month = jun,
       volume = {662},
          eid = {A29},
        pages = {A29},
          doi = {10.1051/0004-6361/202038876},
       adsurl = {https://ui.adsabs.harvard.edu/abs/2022A&A...662A..29M}
}

@ARTICLE{Minor2013,
       author = {{Minor}, Quinn E.},
        title = "{Binary Populations in Milky Way Satellite Galaxies: Constraints from Multi-epoch Data in the Carina, Fornax, Sculptor, and Sextans Dwarf Spheroidal Galaxies}",
      journal = {\apj},
         year = 2013,
        month = dec,
       volume = {779},
       number = {2},
          eid = {116},
        pages = {116},
          doi = {10.1088/0004-637X/779/2/116},
archivePrefix = {arXiv},
       eprint = {1302.0302},
 primaryClass = {astro-ph.GA},
       adsurl = {https://ui.adsabs.harvard.edu/abs/2013ApJ...779..116M}
}

@ARTICLE{Mocz2017,
       author = {{Mocz}, Philip and {Vogelsberger}, Mark and {Robles}, Victor H. and {Zavala}, Jes{\'u}s and {Boylan-Kolchin}, Michael and {Fialkov}, Anastasia and {Hernquist}, Lars},
        title = "{Galaxy formation with BECDM - I. Turbulence and relaxation of idealized haloes}",
      journal = {\mnras},
         year = 2017,
        month = nov,
       volume = {471},
       number = {4},
        pages = {4559-4570},
          doi = {10.1093/mnras/stx1887},
archivePrefix = {arXiv},
       eprint = {1705.05845},
 primaryClass = {astro-ph.CO},
       adsurl = {https://ui.adsabs.harvard.edu/abs/2017MNRAS.471.4559M}
}

@ARTICLE{Moline2017,
       author = {{Molin{\'e}}, {\'A}ngeles and {S{\'a}nchez-Conde}, Miguel A. and {Palomares-Ruiz}, Sergio and {Prada}, Francisco},
        title = "{Characterization of subhalo structural properties and implications for dark matter annihilation signals}",
      journal = {\mnras},
         year = 2017,
        month = apr,
       volume = {466},
       number = {4},
        pages = {4974-4990},
          doi = {10.1093/mnras/stx026},
archivePrefix = {arXiv},
       eprint = {1603.04057},
 primaryClass = {astro-ph.CO},
       adsurl = {https://ui.adsabs.harvard.edu/abs/2017MNRAS.466.4974M}
}

@ARTICLE{Munoz2018,
       author = {{Mu{\~n}oz}, Ricardo R. and {C{\^o}t{\'e}}, Patrick and {Santana}, Felipe A. and {Geha}, Marla and {Simon}, Joshua D. and {Oyarz{\'u}n}, Grecco A. and {Stetson}, Peter B. and {Djorgovski}, S.~G.},
        title = "{A MegaCam Survey of Outer Halo Satellites. III. Photometric and Structural Parameters}",
      journal = {The Astrophysical Journal},
         year = 2018,
        month = jun,
       volume = {860},
       number = {1},
          eid = {66},
        pages = {66},
          doi = {10.3847/1538-4357/aac16b},
archivePrefix = {arXiv},
       eprint = {1806.06891},
 primaryClass = {astro-ph.GA},
       adsurl = {https://ui.adsabs.harvard.edu/abs/2018ApJ...860...66M}
}

@ARTICLE{Nadler2021,
       author = {{Nadler}, E.~O. and {Drlica-Wagner}, A. and {Bechtol}, K. and {Mau}, S. and {Wechsler}, R.~H. and {Gluscevic}, V. and {Boddy}, K. and {Pace}, A.~B. and {Li}, T.~S. and {McNanna}, M. and {Riley}, A.~H. and {Garc{\'\i}a-Bellido}, J. and {Mao}, Y.-Y. and {Green}, G. and {Burke}, D.~L. and {Peter}, A. and {Jain}, B. and {Abbott}, T.~M.~C. and {Aguena}, M. and {Allam}, S. and {Annis}, J. and {Avila}, S. and {Brooks}, D. and {Carrasco Kind}, M. and {Carretero}, J. and {Costanzi}, M. and {da Costa}, L.~N. and {De Vicente}, J. and {Desai}, S. and {Diehl}, H.~T. and {Doel}, P. and {Everett}, S. and {Evrard}, A.~E. and {Flaugher}, B. and {Frieman}, J. and {Gerdes}, D.~W. and {Gruen}, D. and {Gruendl}, R.~A. and {Gschwend}, J. and {Gutierrez}, G. and {Hinton}, S.~R. and {Honscheid}, K. and {Huterer}, D. and {James}, D.~J. and {Krause}, E. and {Kuehn}, K. and {Kuropatkin}, N. and {Lahav}, O. and {Maia}, M.~A.~G. and {Marshall}, J.~L. and {Menanteau}, F. and {Miquel}, R. and {Palmese}, A. and {Paz-Chinch{\'o}n}, F. and {Plazas}, A.~A. and {Romer}, A.~K. and {Sanchez}, E. and {Scarpine}, V. and {Serrano}, S. and {Sevilla-Noarbe}, I. and {Smith}, M. and {Soares-Santos}, M. and {Suchyta}, E. and {Swanson}, M.~E.~C. and {Tarle}, G. and {Tucker}, D.~L. and {Walker}, A.~R. and {Wester}, W. and {DES Collaboration}},
        title = "{Constraints on Dark Matter Properties from Observations of Milky Way Satellite Galaxies}",
      journal = {\prl},
         year = 2021,
        month = mar,
       volume = {126},
       number = {9},
          eid = {091101},
        pages = {091101},
          doi = {10.1103/PhysRevLett.126.091101},
archivePrefix = {arXiv},
       eprint = {2008.00022},
 primaryClass = {astro-ph.CO},
       adsurl = {https://ui.adsabs.harvard.edu/abs/2021PhRvL.126i1101N}
}

@ARTICLE{Navarro1997,
       author = {{Navarro}, Julio F. and {Frenk}, Carlos S. and {White}, Simon D.~M.},
        title = "{A Universal Density Profile from Hierarchical Clustering}",
      journal = {The Astrophysical Journal},
         year = 1997,
        month = dec,
       volume = {490},
       number = {2},
        pages = {493-508},
          doi = {10.1086/304888},
archivePrefix = {arXiv},
       eprint = {astro-ph/9611107},
 primaryClass = {astro-ph},
       adsurl = {https://ui.adsabs.harvard.edu/abs/1997ApJ...490..493N}
}

@ARTICLE{Nori2021,
       author = {{Nori}, Matteo and {Baldi}, Marco},
        title = "{Scaling relations of fuzzy dark matter haloes - I. Individual systems in their cosmological environment}",
      journal = {\mnras},
         year = 2021,
        month = feb,
       volume = {501},
       number = {1},
        pages = {1539-1556},
          doi = {10.1093/mnras/staa3772},
archivePrefix = {arXiv},
       eprint = {2007.01316},
 primaryClass = {astro-ph.CO},
       adsurl = {https://ui.adsabs.harvard.edu/abs/2021MNRAS.501.1539N}
}

@ARTICLE{Pianta2022,
       author = {{Pianta}, Camilla and {Capuzzo-Dolcetta}, Roberto and {Carraro}, Giovanni},
        title = "{The Impact of Binaries on the Dynamical Mass Estimate of Dwarf Galaxies}",
      journal = {\apj},
         year = 2022,
        month = nov,
       volume = {939},
       number = {1},
          eid = {3},
        pages = {3},
          doi = {10.3847/1538-4357/ac9303},
archivePrefix = {arXiv},
       eprint = {2209.08296},
 primaryClass = {astro-ph.GA},
       adsurl = {https://ui.adsabs.harvard.edu/abs/2022ApJ...939....3P}
}

@ARTICLE{Pietrzynski2008,
       author = {{Pietrzy{\'n}ski}, Grzegorz and {Gieren}, Wolfgang and {Szewczyk}, Olaf and {Walker}, Alistair and {Rizzi}, Luca and {Bresolin}, Fabio and {Kudritzki}, Rolf-Peter and {Nalewajko}, Krzysztof and {Storm}, Jesper and {Dall'Ora}, Massimo and {Ivanov}, Valentin},
        title = "{The Araucaria Project: the Distance to the Sculptor Dwarf Spheroidal Galaxy from Infrared Photometry of RR Lyrae Stars}",
      journal = {The Astronomical Journal},
         year = 2008,
        month = jun,
       volume = {135},
       number = {6},
        pages = {1993-1997},
          doi = {10.1088/0004-6256/135/6/1993},
archivePrefix = {arXiv},
       eprint = {0804.0347},
 primaryClass = {astro-ph},
       adsurl = {https://ui.adsabs.harvard.edu/abs/2008AJ....135.1993P}
}

@ARTICLE{Plummer1911,
       author = {{Plummer}, H.~C.},
        title = "{On the problem of distribution in globular star clusters}",
      journal = {\mnras},
         year = 1911,
        month = mar,
       volume = {71},
        pages = {460-470},
          doi = {10.1093/mnras/71.5.460},
       adsurl = {https://ui.adsabs.harvard.edu/abs/1911MNRAS..71..460P}
}

@ARTICLE{Pietrzynski2009,
       author = {{Pietrzy{\'n}ski}, Grzegorz and {G{\'o}rski}, Marek and {Gieren}, Wolfgang and {Ivanov}, Valentin D. and {Bresolin}, Fabio and {Kudritzki}, Rolf-Peter},
        title = "{The Araucaria Project. Infrared Tip of the Red Giant Branch Distances to the Carina and Fornax Dwarf Spheroidal Galaxies}",
      journal = {\aj},
         year = 2009,
        month = aug,
       volume = {138},
       number = {2},
        pages = {459-465},
          doi = {10.1088/0004-6256/138/2/459},
archivePrefix = {arXiv},
       eprint = {0906.0082},
 primaryClass = {astro-ph.GA},
       adsurl = {https://ui.adsabs.harvard.edu/abs/2009AJ....138..459P}
}

@ARTICLE{Prada2012,
       author = {{Prada}, Francisco and {Klypin}, Anatoly A. and {Cuesta}, Antonio J. and {Betancort-Rijo}, Juan E. and {Primack}, Joel},
        title = "{Halo concentrations in the standard {\ensuremath{\Lambda}} cold dark matter cosmology}",
      journal = {\mnras},
         year = 2012,
        month = jul,
       volume = {423},
       number = {4},
        pages = {3018-3030},
          doi = {10.1111/j.1365-2966.2012.21007.x},
archivePrefix = {arXiv},
       eprint = {1104.5130},
 primaryClass = {astro-ph.CO},
       adsurl = {https://ui.adsabs.harvard.edu/abs/2012MNRAS.423.3018P}
}

@ARTICLE{Read2021,
       author = {{Read}, J.~I. and {Mamon}, G.~A. and {Vasiliev}, E. and {Watkins}, L.~L. and {Walker}, M.~G. and {Pe{\~n}arrubia}, J. and {Wilkinson}, M. and {Dehnen}, W. and {Das}, P.},
        title = "{Breaking beta: a comparison of mass modelling methods for spherical systems}",
      journal = {\mnras},
         year = 2021,
        month = feb,
       volume = {501},
       number = {1},
        pages = {978-993},
          doi = {10.1093/mnras/staa3663},
archivePrefix = {arXiv},
       eprint = {2011.09493},
 primaryClass = {astro-ph.GA},
       adsurl = {https://ui.adsabs.harvard.edu/abs/2021MNRAS.501..978R}
}

@ARTICLE{Robles2012,
       author = {{Robles}, Victor H. and {Matos}, T.},
        title = "{Flat central density profile and constant dark matter surface density in galaxies from scalar field dark matter}",
      journal = {\mnras},
         year = 2012,
        month = may,
       volume = {422},
       number = {1},
        pages = {282-289},
          doi = {10.1111/j.1365-2966.2012.20603.x},
archivePrefix = {arXiv},
       eprint = {1201.3032},
 primaryClass = {astro-ph.CO},
       adsurl = {https://ui.adsabs.harvard.edu/abs/2012MNRAS.422..282R}
}

@ARTICLE{Robles2024,
       author = {{Robles}, Victor H. and {Zagorac}, J. Luna and {Padmanabhan}, Nikhil},
        title = "{Scalar field dark matter: impact of supernova-driven blowouts on the soliton structure of low-mass dark matter haloes}",
      journal = {\mnras},
         year = 2024,
        month = aug,
       volume = {532},
       number = {2},
        pages = {1980-1990},
          doi = {10.1093/mnras/stae1544},
archivePrefix = {arXiv},
       eprint = {2308.14691},
 primaryClass = {astro-ph.GA},
       adsurl = {https://ui.adsabs.harvard.edu/abs/2024MNRAS.532.1980R}
}

@ARTICLE{RogersPeiris2021,
       author = {{Rogers}, Keir K. and {Peiris}, Hiranya V.},
        title = "{Strong Bound on Canonical Ultralight Axion Dark Matter from the Lyman-Alpha Forest}",
      journal = {Physical Review Letters},
         year = 2021,
        month = feb,
       volume = {126},
       number = {7},
          eid = {071302},
        pages = {071302},
          doi = {10.1103/PhysRevLett.126.071302},
archivePrefix = {arXiv},
       eprint = {2007.12705},
 primaryClass = {astro-ph.CO},
       adsurl = {https://ui.adsabs.harvard.edu/abs/2021PhRvL.126g1302R}
}

@ARTICLE{Safarzadeh2020,
       author = {{Safarzadeh}, Mohammadtaher and {Spergel}, David N.},
        title = "{Ultra-light Dark Matter Is Incompatible with the Milky Way's Dwarf Satellites}",
      journal = {\apj},
         year = 2020,
        month = apr,
       volume = {893},
       number = {1},
          eid = {21},
        pages = {21},
          doi = {10.3847/1538-4357/ab7db2},
archivePrefix = {arXiv},
       eprint = {1906.11848},
 primaryClass = {astro-ph.CO},
       adsurl = {https://ui.adsabs.harvard.edu/abs/2020ApJ...893...21S}
}

@ARTICLE{SandersEvans2020,
       author = {{Sanders}, Jason L. and {Evans}, N. Wyn},
        title = "{Near-Gaussian distributions for modelling discrete stellar velocity data with heteroskedastic uncertainties}",
      journal = {\mnras},
         year = 2020,
        month = dec,
       volume = {499},
       number = {4},
        pages = {5806-5825},
          doi = {10.1093/mnras/staa2860},
archivePrefix = {arXiv},
       eprint = {2009.07858},
 primaryClass = {astro-ph.GA},
       adsurl = {https://ui.adsabs.harvard.edu/abs/2020MNRAS.499.5806S}
}

@ARTICLE{Schive2014a,
       author = {{Schive}, Hsi-Yu and {Chiueh}, Tzihong and {Broadhurst}, Tom},
        title = "{Cosmic structure as the quantum interference of a coherent dark wave}",
      journal = {Nature Physics},
         year = 2014,
        month = jul,
       volume = {10},
       number = {7},
        pages = {496-499},
          doi = {10.1038/nphys2996},
archivePrefix = {arXiv},
       eprint = {1406.6586},
 primaryClass = {astro-ph.GA},
       adsurl = {https://ui.adsabs.harvard.edu/abs/2014NatPh..10..496S}
}

@ARTICLE{Schive2014b,
       author = {{Schive}, Hsi-Yu and {Liao}, Ming-Hsuan and {Woo}, Tak-Pong and {Wong}, Shing-Kwong and {Chiueh}, Tzihong and {Broadhurst}, Tom and {Hwang}, W. -Y. Pauchy},
        title = "{Understanding the Core-Halo Relation of Quantum Wave Dark Matter from 3D Simulations}",
      journal = {\prl},
         year = 2014,
        month = dec,
       volume = {113},
       number = {26},
          eid = {261302},
        pages = {261302},
          doi = {10.1103/PhysRevLett.113.261302},
archivePrefix = {arXiv},
       eprint = {1407.7762},
 primaryClass = {astro-ph.GA},
       adsurl = {https://ui.adsabs.harvard.edu/abs/2014PhRvL.113z1302S}
}

@ARTICLE{Schive2020,
       author = {{Schive}, Hsi-Yu and {Chiueh}, Tzihong and {Broadhurst}, Tom},
        title = "{Soliton Random Walk and the Cluster-Stripping Problem in Ultralight Dark Matter}",
      journal = {\prl},
         year = 2020,
        month = may,
       volume = {124},
       number = {20},
          eid = {201301},
        pages = {201301},
          doi = {10.1103/PhysRevLett.124.201301},
archivePrefix = {arXiv},
       eprint = {1912.09483},
 primaryClass = {astro-ph.GA},
       adsurl = {https://ui.adsabs.harvard.edu/abs/2020PhRvL.124t1301S}
}

@ARTICLE{Schwabe2016,
       author = {{Schwabe}, Bodo and {Niemeyer}, Jens C. and {Engels}, Jan F.},
        title = "{Simulations of solitonic core mergers in ultralight axion dark matter cosmologies}",
      journal = {Physical Review D},
         year = 2016,
        month = aug,
       volume = {94},
       number = {4},
          eid = {043513},
        pages = {043513},
          doi = {10.1103/PhysRevD.94.043513},
archivePrefix = {arXiv},
       eprint = {1606.05151},
 primaryClass = {astro-ph.CO},
       adsurl = {https://ui.adsabs.harvard.edu/abs/2016PhRvD..94d3513S}
}

@ARTICLE{Skuladottir2023,
       author = {{Sk{\'u}lad{\'o}ttir}, {\'A}. and {Puls}, A.~A. and {Amarsi}, A.~M. and {Battaglia}, G. and {Buder}, S. and {Campbell}, S. and {Cardona-Barrero}, S. and {Christlieb}, N. and {Feuillet}, D.~K. and {Gelli}, V. and {Hansen}, C.~J. and {Hill}, V. and {Ibata}, R. and {Jablonka}, P. and {Kacharov}, N. and {Karakas}, A. and {Koch-Hansen}, A.~J. and {Lind}, K. and {Lombardo}, L. and {Lucchesi}, R.~E.~R. and {Lugaro}, M. and {Martin}, N. and {Massari}, D. and {Nordlander}, T. and {Reichert}, M. and {Rossi}, M. and {Ruiter}, A.~J. and {Salvadori}, S. and {Seitenzahl}, I.~R. and {Tolstoy}, E. and {Xylakis-Dornbusch}, T. and {Youakim}, K.~C.},
        title = "{The 4MOST Survey of Dwarf Galaxies and their Stellar Streams (4DWARFS)}",
      journal = {The Messenger},
         year = 2023,
        month = mar,
       volume = {190},
        pages = {19-21},
          doi = {10.18727/0722-6691/5304},
       adsurl = {https://ui.adsabs.harvard.edu/abs/2023Msngr.190...19S}
}

@ARTICLE{Spencer2018,
       author = {{Spencer}, Meghin E. and {Mateo}, Mario and {Olszewski}, Edward W. and {Walker}, Matthew G. and {McConnachie}, Alan W. and {Kirby}, Evan N.},
        title = "{The Binary Fraction of Stars in Dwarf Galaxies: The Cases of Draco and Ursa Minor}",
      journal = {The Astronomical Journal},
         year = 2018,
        month = dec,
       volume = {156},
       number = {6},
          eid = {257},
        pages = {257},
          doi = {10.3847/1538-3881/aae3e4},
archivePrefix = {arXiv},
       eprint = {1811.06597},
 primaryClass = {astro-ph.GA},
       adsurl = {https://ui.adsabs.harvard.edu/abs/2018AJ....156..257S}
}

@ARTICLE{Spencer2017,
       author = {{Spencer}, Meghin E. and {Mateo}, Mario and {Walker}, Matthew G. and {Olszewski}, Edward W.},
        title = "{A Multi-epoch Kinematic Study of the Remote Dwarf Spheroidal Galaxy Leo II}",
      journal = {\apj},
         year = 2017,
        month = feb,
       volume = {836},
       number = {2},
          eid = {202},
        pages = {202},
          doi = {10.3847/1538-4357/836/2/202},
archivePrefix = {arXiv},
       eprint = {1702.08836},
 primaryClass = {astro-ph.GA},
       adsurl = {https://ui.adsabs.harvard.edu/abs/2017ApJ...836..202S}
}

@ARTICLE{Stott2018,
       author = {{Stott}, Matthew J. and {Marsh}, David J.~E.},
        title = "{Black hole spin constraints on the mass spectrum and number of axionlike fields}",
      journal = {\prd},
         year = 2018,
        month = oct,
       volume = {98},
       number = {8},
          eid = {083006},
        pages = {083006},
          doi = {10.1103/PhysRevD.98.083006},
archivePrefix = {arXiv},
       eprint = {1805.02016},
 primaryClass = {hep-ph},
       adsurl = {https://ui.adsabs.harvard.edu/abs/2018PhRvD..98h3006S}
}

@ARTICLE{Takada2014,
       author = {{Takada}, Masahiro and {Ellis}, Richard S. and {Chiba}, Masashi and {Greene}, Jenny E. and {Aihara}, Hiroaki and {Arimoto}, Nobuo and {Bundy}, Kevin and {Cohen}, Judith and {Dor{\'e}}, Olivier and {Graves}, Genevieve and {Gunn}, James E. and {Heckman}, Timothy and {Hirata}, Christopher M. and {Ho}, Paul and {Kneib}, Jean-Paul and {Le F{\`e}vre}, Olivier and {Lin}, Lihwai and {More}, Surhud and {Murayama}, Hitoshi and {Nagao}, Tohru and {Ouchi}, Masami and {Seiffert}, Michael and {Silverman}, John D. and {Sodr{\'e}}, Laerte and {Spergel}, David N. and {Strauss}, Michael A. and {Sugai}, Hajime and {Suto}, Yasushi and {Takami}, Hideki and {Wyse}, Rosemary},
        title = "{Extragalactic science, cosmology, and Galactic archaeology with the Subaru Prime Focus Spectrograph}",
      journal = {Publications of the Astronomical Society of Japan},
         year = 2014,
        month = feb,
       volume = {66},
       number = {1},
          eid = {R1},
        pages = {R1},
          doi = {10.1093/pasj/pst019},
archivePrefix = {arXiv},
       eprint = {1206.0737},
 primaryClass = {astro-ph.CO},
       adsurl = {https://ui.adsabs.harvard.edu/abs/2014PASJ...66R...1T}
}

@INPROCEEDINGS{Tamura2016,
       author = {{Tamura}, Naoyuki and {Takato}, Naruhisa and {Shimono}, Atsushi and {Moritani}, Yuki and {Yabe}, Kiyoto and {Ishizuka}, Yuki and {Ueda}, Akitoshi and {Kamata}, Yukiko and {Aghazarian}, Hrand and {Arnouts}, St{\'e}phane and {Barban}, Gabriel and {Barkhouser}, Robert H. and {Borges}, Renato C. and {Braun}, David F. and {Carr}, Michael A. and {Chabaud}, Pierre-Yves and {Chang}, Yin-Chang and {Chen}, Hsin-Yo and {Chiba}, Masashi and {Chou}, Richard C.~Y. and {Chu}, You-Hua and {Cohen}, Judith and {de Almeida}, Rodrigo P. and {de Oliveira}, Antonio C. and {de Oliveira}, Ligia S. and {Dekany}, Richard G. and {Dohlen}, Kjetil and {dos Santos}, Jesulino B. and {dos Santos}, Leandro H. and {Ellis}, Richard and {Fabricius}, Maximilian and {Ferrand}, Didier and {Ferreira}, D{\'e}cio and {Golebiowski}, Mirek and {Greene}, Jenny E. and {Gross}, Johannes and {Gunn}, James E. and {Hammond}, Randolph and {Harding}, Albert and {Hart}, Murdock and {Heckman}, Timothy M. and {Hirata}, Christopher M. and {Ho}, Paul and {Hope}, Stephen C. and {Hovland}, Larry and {Hsu}, Shu-Fu and {Hu}, Yen-Shan and {Huang}, Ping-Jie and {Jaquet}, Marc and {Jing}, Yipeng and {Karr}, Jennifer and {Kimura}, Masahiko and {King}, Matthew E. and {Komatsu}, Eiichiro and {Le Brun}, Vincent and {Le F{\`e}vre}, Olivier and {Le Fur}, Arnaud and {Le Mignant}, David and {Ling}, Hung-Hsu and {Loomis}, Craig P. and {Lupton}, Robert H. and {Madec}, Fabrice and {Mao}, Peter and {Marrara}, Lucas S. and {Mendes de Oliveira}, Claudia and {Minowa}, Yosuke and {Morantz}, Chaz and {Murayama}, Hitoshi and {Murray}, Graham J. and {Ohyama}, Youichi and {Orndorff}, Joseph and {Pascal}, Sandrine and {Pereira}, Jefferson M. and {Reiley}, Daniel and {Reinecke}, Martin and {Ritter}, Andreas and {Roberts}, Mitsuko and {Schwochert}, Mark A. and {Seiffert}, Michael D. and {Smee}, Stephen A. and {Sodre}, Laerte and {Spergel}, David N. and {Steinkraus}, Aaron J. and {Strauss}, Michael A. and {Surace}, Christian and {Suto}, Yasushi and {Suzuki}, Nao and {Swinbank}, John and {Tait}, Philip J. and {Takada}, Masahiro and {Tamura}, Tomonori and {Tanaka}, Yoko and {Tresse}, Laurence and {Verducci}, Orlando and {Vibert}, Didier and {Vidal}, Clement and {Wang}, Shiang-Yu and {Wen}, Chih-Yi and {Yan}, Chi-Hung and {Yasuda}, Naoki},
        title = "{Prime Focus Spectrograph (PFS) for the Subaru telescope: overview, recent progress, and future perspectives}",
    booktitle = {Ground-based and Airborne Instrumentation for Astronomy VI},
         year = 2016,
       editor = {{Evans}, Christopher J. and {Simard}, Luc and {Takami}, Hideki},
       series = {Society of Photo-Optical Instrumentation Engineers (SPIE) Conference Series},
       volume = {9908},
        month = aug,
          eid = {99081M},
        pages = {99081M},
          doi = {10.1117/12.2232103},
archivePrefix = {arXiv},
       eprint = {1608.01075},
 primaryClass = {astro-ph.IM},
       adsurl = {https://ui.adsabs.harvard.edu/abs/2016SPIE.9908E..1MT}
}
\bibliographystyle{aasjournal}

\end{document}